\documentclass{article}

\usepackage{arxiv}

\usepackage{amsthm}
\usepackage{amsmath}
\usepackage{amsfonts}
\usepackage{amssymb}

\usepackage[utf8]{inputenc} 
\usepackage[T1]{fontenc}    
\usepackage{hyperref}       
\usepackage{url}            
\usepackage{booktabs}       
\usepackage{amsfonts}       
\usepackage{nicefrac}       
\usepackage{microtype}      
\usepackage{cleveref}       
\usepackage{lipsum}         
\usepackage{graphicx}
\usepackage{natbib}
\usepackage{doi}

\usepackage{algorithm2e}
\usepackage{algorithmic}
\usepackage{tikz}
\usetikzlibrary{bayesnet}

\providecommand{\keywords}[1]
{
  \small
  \textbf{\textit{Keywords---}} #1
}

\newcommand{\mP}{\mathcal P}

\usepackage{todonotes}

\usepackage{thm-restate}

\title{A probabilistic model of ocean floats under ice}

\author{Derek Hansen\thanks{Corresponding author.} \\
	Department of Statistics\\
	University of Michigan\\
	\texttt{dereklh@umich.edu} \\
	\And
    Drew Yarger\\
	Department of Statistics\\
	University of Michigan\\
	\texttt{dyarger@umich.edu} \\
}

\renewcommand{\shorttitle}{ArgoSSM}

\hypersetup{
pdftitle={ArgoSSM: A State-space Model of Ocean Floats Under Ice},
pdfsubject={q-bio.NC, q-bio.QM},
pdfauthor={Derek Hansen, Drew Yarger},
pdfkeywords={Sequential Monte Carlo, Kalman filter, physical oceanography, sensor network, spatiotemporal modeling},
}

\begin{document}
\maketitle

\begin{abstract}
	
The Argo project deploys thousands of floats throughout the world’s oceans. Carried only by the current, these floats take measurements such as temperature and salinity at depths of up to two kilometers. These measurements are critical for scientific tasks such as modeling climate change, estimating temperature and salinity fields, and tracking the global hydrological cycle. In the Southern Ocean, Argo floats frequently drift under ice cover which prevents tracking via GPS. Managing this missing location data is an important scientific challenge for the Argo project. To predict the floats’ trajectories under ice and quantify their uncertainty, we introduce a probabilistic state-space model (SSM) called ArgoSSM. ArgoSSM infers the posterior distribution of a float’s position and velocity at each time based on all available data, which includes GPS measurements, ice cover, and potential vorticity. This inference is achieved via an efficient particle filtering scheme, which is effective despite the high signal-to-noise ratio in the GPS data. Compared to existing interpolation approaches in oceanography, ArgoSSM more accurately predicts held-out GPS measurements. Moreover, because uncertainty estimates are well-calibrated in the posterior distribution, ArgoSSM enables more robust and accurate temperature, salinity, and circulation estimates.

\end{abstract}

\keywords={Sequential Monte Carlo, Kalman filter, physical oceanography, sensor network, spatiotemporal modeling}

\section{Introduction}

Reliable measurements of the ocean are essential for scientific tasks such as modeling climate change \citep{lyman_estimating_2014}, estimating temperature and salinity fields \citep{chang_objective_2009}, and tracking the global hydrological cycle \citep{hosodaGlobalSurfaceLayer2009}.
However, measuring the vast reaches of the ocean is a challenging task.
Historically, measurements were taken by sensors onboard ships, but this limited most collection to popular routes, far from a comprehensive survey of the ocean.

To address this issue, the Argo project \citep{argo2020} was started in 1998 as a multinational collaboration to collect data about the world's oceans.
Instead of using ships, the Argo project deploys floats that freely drift with the ocean's currents at a depth of one kilometer.
Every ten days, a float descends to two kilometers and then ascends to the surface, measuring temperature, salinity, and pressure at multiple depths along the way.
At the surface, the float records its position via GPS and transmits data via the Iridium satellite system to Argo data centers for processing.
Each iteration of this data collection process is called a profile.

\begin{figure}[t]
  \centering
\includegraphics[width=.65\linewidth]{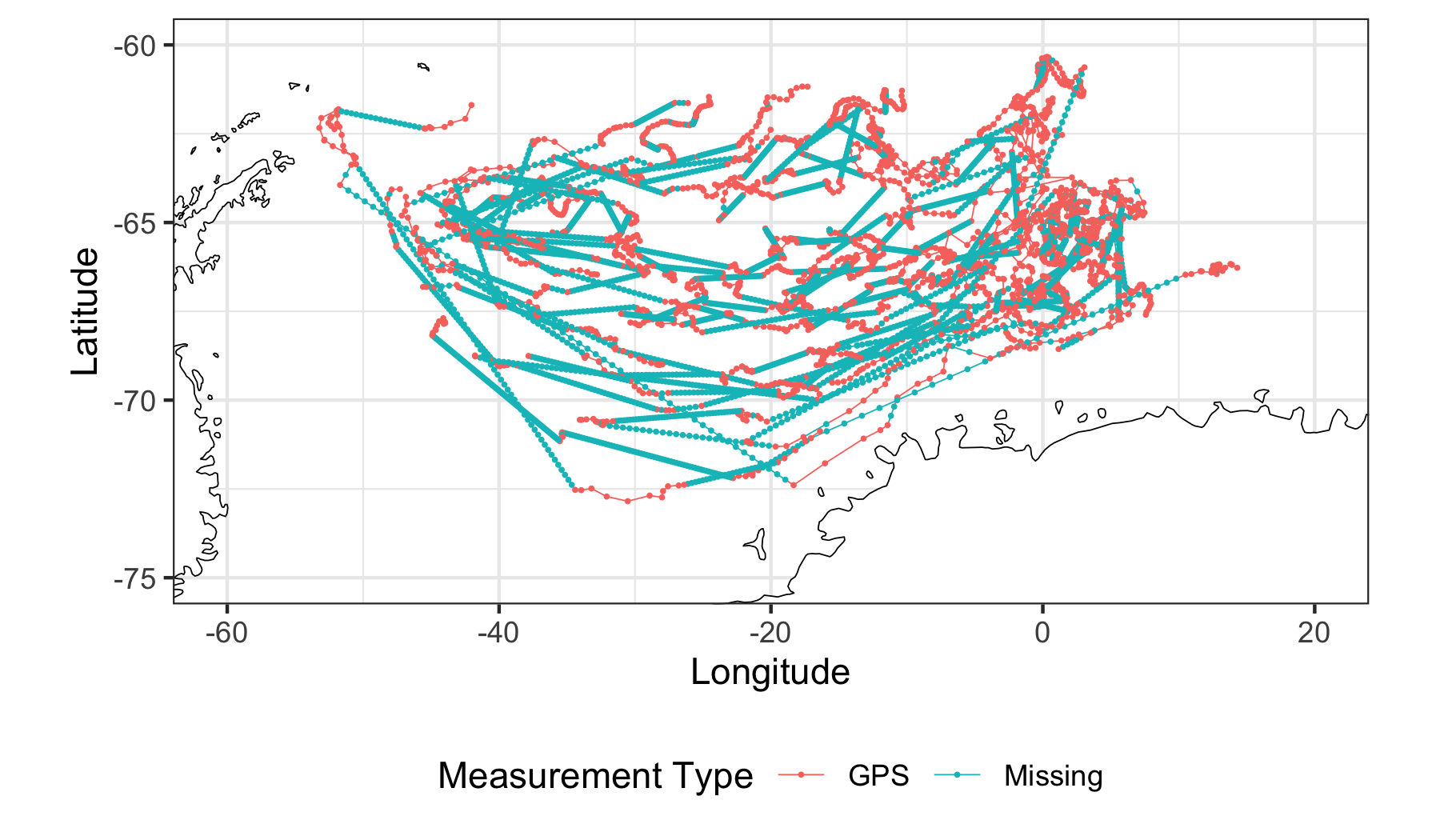}
\includegraphics[width=.3\linewidth]{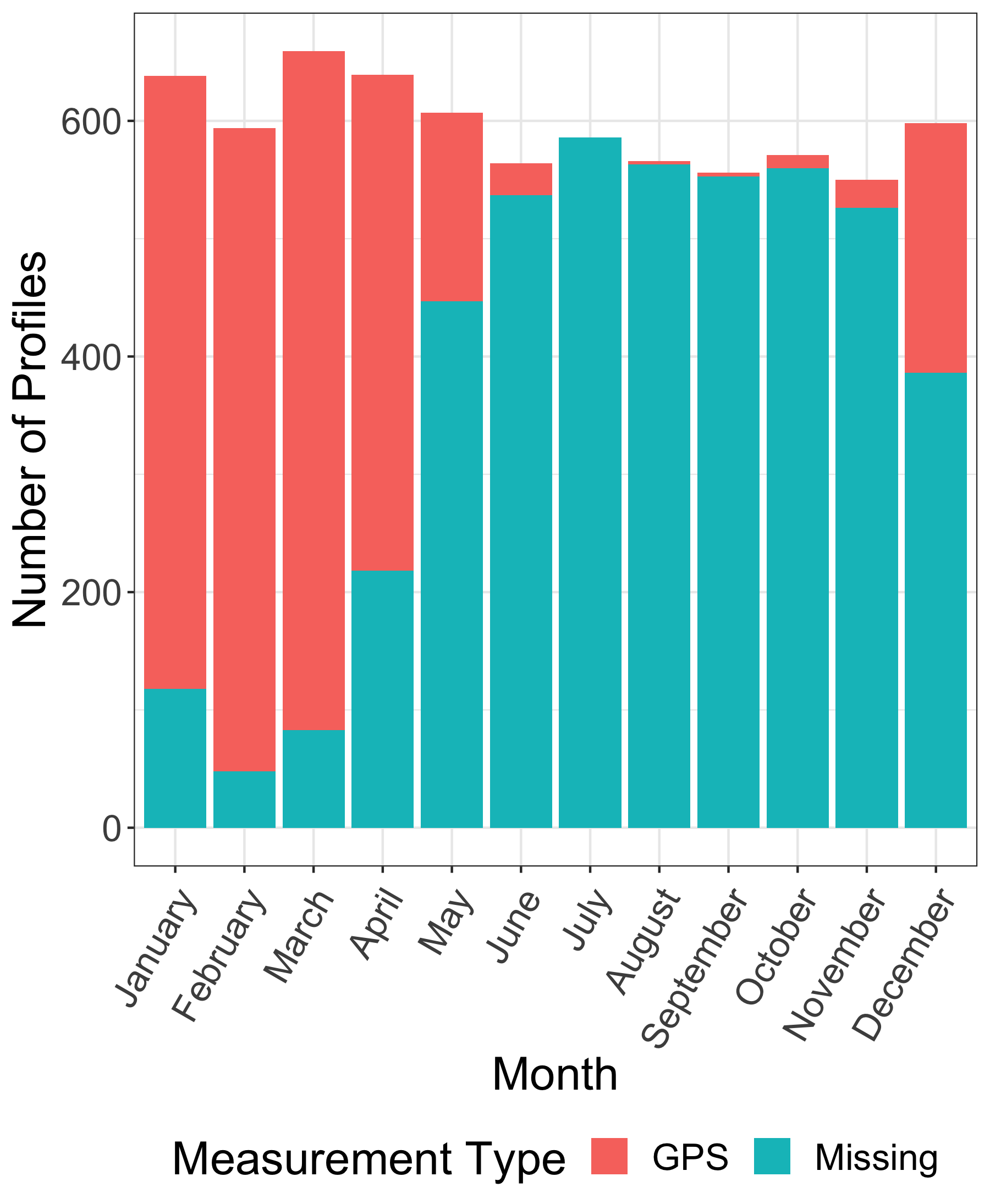}
  \caption{(Left) Recorded locations of 46 selected floats in the Weddell Gyre, a region of the Southern Ocean. Each dot represents a profile, a collection of temperature and salinity measurements. In this plot, missing locations of profiles are linearly interpolated between valid location measurements and do not reflect the floats' true positions (or the positions' uncertainties). (Right) Number of profiles collected by these floats in each month of the year. Nearly all profiles collected in the winter had missing locations.}
  \label{fig:floatoverview}
\end{figure}

The implementation of an ice avoidance program in \citet{klatt_profiling_2007} has enabled floats to explore previously inaccessible regions such as the Southern Ocean near Antarctica.
In these regions, when ice cover is detected, an Argo float does not attempt to resurface to avoid damage.
While the float continues to take profiles as normal, GPS tracking can be lost for more than six months.
Because the float moves freely, GPS tracking is critical for pinpointing its whereabouts when data is collected.
Figure~\ref{fig:floatoverview} illustrates both the extent and seasonality of missing locations in the Southern Ocean.
Profiles with missing locations make up $16\%$ of the Argo dataset in the Southern Ocean \citep{chamberlain_observing_2018, reeve_horizontal_2019}.
Thus, simply removing the measurements without locations leads to both spatial and seasonal biases in estimation tasks \citep{gray_global_2014, reeve_gridded_2016, reeve_horizontal_2019}.
In turn, these biases can affect our understanding of systems in which the Southern Ocean plays an important role, such as the global climate \citep[e.g.][]{gray_autonomous_2018}.

The most common way to handle missing locations is to linearly interpolate between the nearest two observed locations \citep{wong_argo_2020}.
\citet{chamberlain_observing_2018} improved this technique by interpolating a path based on local estimates of potential vorticity (PV), which takes into account the effect of the Earth's rotation on the ocean water column \citep[see Section 7.7 of][for more information]{talley_chapter_2011}.
In parallel, \citet{chamberlain_observing_2018} showed how positioning data from sound sources could constrain the locations using a velocity-driven linear state-space model (SSM), but with fixed parameters and no PV component.
While their underlying model is probabilistic, they only use the predicted mean and do not report any model-estimated uncertainty.

All of these previous approaches fail to offer well-calibrated estimates of uncertainty alongside their predictions.
\citet{chamberlain_observing_2018} estimated that location uncertainty can be as much as $116\text{km}$. However, this estimate came from the error in held-out data estimates and is not available for unseen data points actually under the ice.
Moreover, these approaches do not offer a way to incorporate location uncertainty into downstream estimation tasks, leading to overconfident final estimates.

To solve both of these issues, we introduce a probabilistic state-space model of Argo float movements.
Our model, ArgoSSM, combines available GPS measurements with local conservation of potential vorticity (PV) to more accurately model the floats' positions and velocities.
In addition, we use data on ice concentration in the Southern Ocean to further constrain the floats' positions and directly characterize the missing data mechanism in our statistical model.
Because ArgoSSM is a generative model, we infer a posterior distribution of the trajectory of each float as well as model parameters given all available information.
This inference is achieved via an efficient particle filtering scheme that fully adapts to the high signal-to-noise ratio in the GPS data.
In a case study of Argo floats in the Southern Ocean, ArgoSSM more accurately predicts held-out GPS measurements than previous interpolation methods.
Moreover, ArgoSSM illuminates where location data is particularly sparse and uncertain, quantifying a source of uncertainty for downstream estimates (e.g., temperature, salinity, or ocean circulation) that would have been ignored with imputed location estimates.
By providing a principled and flexible statistical framework to handle missing location measurements, ArgoSSM can easily be incorporated into future scientific study of the Southern Ocean.

\section{A generative model of ocean float movement}\label{sec:state-space-model}

To motivate the probabilistic framework of ArgoSSM, we start with a simple model of how Argo floats move through the ocean.
Profile data is collected at times $t_1 < t_2 < \dots t_N$, approximately ten days apart.
At each time $t_n$, let $X_n$ be the geographic position in latitude and longitude at time $t_n$.
We take into account the elapsed time $\Delta t_n = t_n - t_{n-1}$ when updating the position from $X_{n-1}$ to $X_n$.
If the float's position at $n-1$ was $X_{n-1}$, we might expect that $X_{n}$ will be close to $X_{n-1}$ with noise proportional to the time passed.
This can be written explicitly as a two-dimensional random walk (RW) model:
\begin{equation}  \label{eq:1}
      X_{n} = X_{n-1} + \epsilon^{X}_{n},
\end{equation}
where $\epsilon_n^X$ follows a multivariate Gaussian distribution with zero mean and covariance $\Delta t_n \Sigma_X$.

For a given index $n$, the expected value of $X_{n}$ conditioned on $X_{n-1}$ and $X_{n+1}$ is a time-weighted average of $X_{n-1}$ and $X_{n+1}$.
Thus, the RW model is a generative model of float movement where the optimal predictor for an unseen point is linear interpolation.
Linear interpolation might work well for short gaps in time, but it breaks down for large gaps in time that are seen in the Southern Ocean.
This because linear interpolation ignores local information such as momentum.
If we know the current has carried the float from position $X_{n-1}$ to position $X_{n}$, that same current will likely carry the float further in the same direction.
More specifically, each float has a velocity $V_{n}$ that indicates where it is headed next.
Thus, we modify Equation \ref{eq:1} to take velocity into account:
\begin{equation}
  \label{eq:2}
  X_{n} = X_{n-1} + \Delta t_n V_{n-1} + \epsilon^{X}_{n} .
\end{equation}
The velocity $V_{n}$ also changes over time according to an auto-regressive (AR) model:
\begin{equation}\label{eq:3}
  V_{n} = (1 - \alpha^{\Delta t_n}) v_{0} + \alpha^{\Delta t_n} V_{n-1} + \epsilon^{V}_{n},
\end{equation}
where $\epsilon_n^V$ follows a multivariate Gaussian distribution with zero mean and covariance $\Delta t_n \Sigma_V$.
Two parameters govern the velocity update: the long-run velocity of the float $v_0$ and the autoregressive term $\alpha \in [0, 1]$.
The parameter $\alpha$ determines how quickly the velocity reverts to the long-run velocity $v_0$.

We refer to Equations \ref{eq:2} and \ref{eq:3} collectively as the AR model.
While the AR model is more realistic than the RW model, it simplifies the true behavior of the floats.
Notably, it ignores the float's vertical movement as it rises or drops in the ocean and intra-day movement on the ocean surface.
Thus, the velocity state-variable should be interpreted as the average direction of the float over several days rather than a local estimate of the instantaneous velocity.
The AR model is similar to the state-space model introduced in \citet{chamberlain_observing_2018}, though with key differences.
The main modeling difference is that \citet{chamberlain_observing_2018} updates the velocity according to a random walk (corresponding to $\alpha=1$ in Equation~\ref{eq:3}).
In Section~\ref{sec:experiments}, we find for many floats that the inferred autoregressive parameter $\alpha$ is significantly less than $1$.
Moreover, \citet{chamberlain_observing_2018} fixes all parameter values, whereas we estimate them alongside the positions and velocities.

Information about the float's position comes from GPS measurements $Y_{n}$ corresponding to each $X_{n}$:
\begin{equation}
  \label{eq:5}
Y_{n} = X_{n} + \epsilon^{Y}_{n},
\end{equation}
where $\epsilon^Y_n$ is the measurement error that follows a multivariate Gaussian with zero mean and covariance $\Sigma_Y$.
With the Iridium satellite system, Argo GPS measurements are rated to be accurate to within eight meters \citep{wong_argo_2020}, so the variability of the measurement error $\epsilon_n^Y$ in Equation \ref{eq:5} will typically be magnitudes lower than that of the transition error $\epsilon_n^X$ in Equation \ref{eq:2}.

\subsection{Missing due to ice cover}

While GPS measurements accurately pin down the floats' locations, they may not always be available.
To represent this availability, let $A_n$ be an indicator variable that equals one if GPS is available at time $t_n$ and zero otherwise.
In the Southern Ocean, since the float only surfaces after three consecutive ice-free detections, $A_n$ is mostly determined by the ice-avoidance algorithm \citep{klatt_profiling_2007}, which depends on the concentration of ice in the area.

To model the availability indicator $A_n$, we first require an estimated probability of detecting ice.
We have available daily ice concentration estimates from \citet{ice_data}, which uses remotely-sensed data from microwave instruments on satellites.
Let $E(x, t)$ be the concentration of ice at position $x$ and time $t$.
Accounting for imperfect ice detection due to limited resolution, the probability that the float detects ice at position $x$ and time $t$ is
\(
\tilde E(x, t) = p^{\text{TPR}} E(x, t) + (1 - p^{\text{TNR}}) E(x, t)\),
where $p^{\text{TPR}}$ is the ``true positive rate'' (correctly detecting ice) and $p^{\text{TNR}}$ is the ``true negative rate'' (correctly detecting no ice).
We expect $p^{\text{TPR}}$ and $p^{\text{TNR}}$ to be close to $1$, but since detections are based on the temperature of the water, we expect to see more false positives than false negatives (i.e. $p^{\text{TNR}} < p^{\text{TPR}}$).

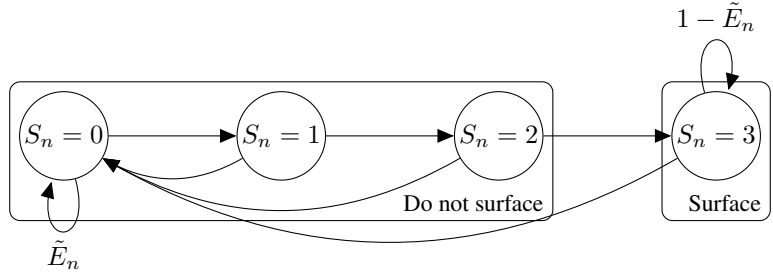
\begin{figure}[t]
	\begin{center}
	   \begin{tikzpicture}[x=1.7cm,y=0.25cm]

		\node[latent] (state0) {$S_n = 0$} ; %
		\node[latent, right=of state0] (state1) {$S_n = 1$} ; %
		\node[latent, right=of state1] (state2) {$S_n = 2$} ; %
		\node[latent, right=of state2] (state3) {$S_n = 3$} ; %

		\edge {state0} {state1};
		\edge {state1} {state2};
		\edge {state2} {state3};
		\path (state3) edge [->, >={triangle 45}, loop above] node (state3') {$1 - \tilde E_n$} (state3);

		\edge [bend left] {state3} {state0};
		\edge [bend left] {state2} {state0};
		\edge [bend left] {state1} {state0};
		\path (state0) edge [->, >={triangle 45}, loop below] node (state0') {$\tilde E_n$} (state0);

		\plate {plate1} {
		  (state0)(state1)(state2)
		} {Do not surface}

		\plate {plate2} {
		  (state3)
		} {Surface}

	   \end{tikzpicture}
	\end{center}
    \caption{Transition diagram of the state $S_{n}$ of the Argo ice detection algorithm. The transition to the next state is determined based on whether ice is detected or not.
    The probability of detecting ice, $\tilde E_n \equiv \tilde E(X_n, t_n)$, depends on both the geographic position $X_n$ and the time $t_n$.  Surfacing requires at least three consecutive ice-free detections.} \label{fig:ice_detection_algo}
  \end{figure}

With the probability of detecting ice at a particular time and location, we model the state of the ice-avoidance algorithm as a Markov chain, illustrated in \autoref{fig:ice_detection_algo}.
The state of the algorithm, denoted $S_n$, can take one of four possible values in $\{0, 1, 2, 3\}$.
If ice is detected, the state $S_n$ resets to $0$.
Otherwise, the state increments by one (i.e. $S_n = (S_{n-1} + 1) \wedge 3$).
The probability of transitioning to $S_n=0$ is equal to the probability of detecting ice $\tilde E(X_n, t_n)$.
Likewise, the probability of maintaining the streak of ice-free detections is $1 - \tilde E(X_n, t_n)$.
The ice-avoidance state $S_n$ determines whether the float surfaces, which directly impacts the availability of the GPS measurement $A_n$.
For $S_n \in \{0, 1, 2\}$, the float will not surface, so the measurement is missing ($A_n = 0$) with probability 1.
If $S_n = 3$, the ice-avoidance algorithm will no longer prevent the float from surfacing.
In this case, $P(A_n = 1 | S_n = 3) = (1 - p^{\text{MAR}})$,
where $p^{\text{MAR}}$ is the probability that the GPS measurement is missing for reasons other than ice avoidance.

Even if not of direct interest, knowledge of the ice-avoidance state $S_n$ provides information about the float's position.
In particular, whenever $S_n$ equals $0$, there was most likely ice present at position $X_n$, so it is more likely the float is in a region with high concentration of ice.
Similarly, if $S_n \in \{1, 2, 3\}$, then the float did not detect ice, so it is more likely the float is in a region with a low concentration of ice.
This relationship is naturally captured in the joint posterior distribution of the three state variables ($X_n, V_n, S_n$) given the observed data. We discuss how this posterior distribution is estimated in Section \ref{sec:smc_proposal}.
\subsection{Local conservation of potential vorticity}
\newcommand{\pvfn}{\mathrm{PV}}
\newcommand{\pv}{\nabla {\pvfn}{(X_n)}}

The motion of a freely-circulating object in the ocean conserves potential vorticity (PV) \citep{talley_chapter_2011}.
PV is a function of the depth of the ocean and the vorticity, or local spin, which itself is a function of latitude.
Incorporating PV conservation is important for both generating more realistic float trajectories and improving predictive performance in periods without GPS measurements.
To incorporate local conservation of PV into the state-space model, we frame it as a probabilistic constraint.
From a first-order Taylor approximation and Equation \ref{eq:2}, we have that the difference in PV between two positions $X_{n+1}$ and $X_{n}$ is approximately $\epsilon_n^{\text{PV}} = \pv \cdot \Delta t_n V_n$, where $\pv$ is the gradient of PV with respect to position.
Since PV is a conserved quantity, $\epsilon_n^{\text{PV}}$ should be close to zero, but not exactly zero due to the first-order approximation and imperfect estimation of PV.
To account for this, we suppose $\epsilon_n^{\text{PV}}$ follows a univariate Gaussian distribution with mean $0$ and standard deviation $\Delta t_n \sigma_{PV}$, with $\sigma_{PV}$ determining the relative strictness of PV conservation.

Because $\epsilon_n^{\text{PV}}$ is linear with respect to $V_n$ and Gaussian, it is an implicit measurement of $V_n$ that can be incorporated as a Bayesian update of the autoregressive velocity update from Equation \ref{eq:3}:
\begin{equation} \label{eq:pv_posterior}
V_n \vert X_n, V_{n - 1} = B\left((1 - \alpha^{\Delta t_n}) v_{0} + \alpha^{\Delta t_n} V_{n-1} \right) + B^{\frac 1 2}\epsilon^{V}_{n},
\end{equation}
where $B = \left(I + \frac 1 {\sigma_{PV}^2} (\Delta t_n \Sigma_V)  \pv \pv^\prime\right)^{-1}$ is a matrix that encompasses the effect of PV conservation.
The form of $B$ follows from a conjugate Bayesian update with a Gaussian prior and likelihood \cite[see Section 3.5 of][]{gelmanBayesianDataAnalysis2015}.
Notably, $B$ shrinks the component of velocity in the direction of $\pv$ while leaving the other component unchanged.
This discourages large changes in PV, leading to
more realistic predictions of under ice float trajectories.
In turn, this improves predictive performance during periods with no GPS measurements (Section \ref{sec:experiments}).

The value of $\nabla \textrm{PV}(X_n)$ used in the update is estimated from known ocean depth and latitude. 
We use bathymetry exported from the Southern Ocean State Estimate (SOSE) \citep{verdy_data_2017}. 
Potential vorticity is approximated by the expression $\textrm{PV}(X_n) = f(X_n)/h(X_n)$ where $h(X_n)$ is the ocean depth at $X_n$ and $f(X_n) =2\Omega \sin\left(\pi X_{n,lat}/180\right)$ is the Coriolis parameter based on the latitude of $X_n$ and $\Omega = 7.292115 \cdot 10^{-5}$.
See \cite{talley_chapter_2011} for more specifics on how PV is defined and its interpretation.

As in \cite{chamberlain_observing_2018}, we smooth the bathymetry to improve the results, and furthermore we estimate the gradient of PV using local quadratic regression \citep{fan_local_1996}.
We use a bandwidth of $300$ kilometers for local regression, though other bandwidths can easily be integrated into our analysis pipeline.
These regressions result in the smoothed estimates $\hat{h}(\cdot)$ of ocean depth and $\nabla \hat{h}(\cdot)$ of its gradient. 
With these values, the estimate of the PV gradient is \begin{equation}\nabla \textrm{PV}(X_n) = \frac{2\pi \Omega\cos\left(\frac{\pi X_{n,lat}}{180}\right)}{180 \hat{h}(X_n)}\begin{pmatrix} 0 \\ 1 \end{pmatrix} - \frac{f(X_n)}{\hat{h}(X_n)^2} \nabla \hat{h}(X_n)\end{equation}
from the quotient derivative rule.
These estimates are precomputed on a fine grid of locations, then bilinearly interpolated for efficient inference.

\section{State and parameter estimation with sequential Monte Carlo}\label{sec:smc_proposal}

As a state-space model, the distribution of float trajectories defined by ArgoSSM is composed of three parts:
an initial distribution $\mu_{\theta}(X_{1}, V_{1}, S_1)$ in the float's initial state, a conditional transition distribution from the previous state to the current state $f_{\theta} (X_{n}, V_{n}, S_{n} | X_{n-1}, V_{n-1}, S_{n-1})$, and a measurement distribution $g_{\theta}(A_n, Y_n | S_n, X_n)$.
The parameter $\theta$ encapsulates the parameters of each equation.
The measurement distribution can be further decomposed into whether GPS is available $g_{\theta}^A (A_n | S_n)$ and the distribution of the GPS reading $g_{\theta}^Y (Y_{n} | X_{n}) $ when $A_n = 1$.
This allows the distribution of the positions, velocities, and available measurements to factorize as follows:
\begin{equation}\label{eq:state-space-model}
 \begin{split}
     \mP_{\theta} (X_{1:N}, V_{1:N}, S_{1:N}, A_{1:N}, Y_{n: A_n = 1}) = \mu_{\theta}(X_1, V_1, S_1) &\prod_{n=2}^{N} f_{\theta} (X_{n}, V_{n}, S_n| X_{{n-1}} , V_{{n-1}}, S_{n-1})\\ &\prod_{n=2}^N g_{\theta}^A (A_n | S_n) \prod_{A_n = 1} g_{\theta}^Y (Y_n | X_{{n}})
 \end{split}
\end{equation}
For the RW and AR models, the initial, transition, and measurement distributions are Gaussian, the expected values of the transition and measurement distributions are affine functions of their dependencies, and the missingness indicator $A_n$ is assumed to be independent of $X_{n}$ and $Y_{n}$.
This means that both $A_{1:N}$ and $S_{1:N}$ can be integrated out of Equation \ref{eq:state-space-model} above, leaving the marginal distribution of the remaining variables as a multivariate Gaussian with a sparse covariance matrix.
Thus, sampling from the posterior $\mP_\theta (X_{1:N}, V_{1:N} | Y)$ is tractable and efficient via a Kalman smoother.
See \citet{lopesParticleFiltersBayesian2011} for an overview of Kalman filtering and smoothing.

Incorporating ice cover information and local conservation of PV introduces non-linearities that make the Kalman filter inapplicable.
Instead, to sample from the posterior, we use Sequential Monte Carlo (SMC), or a particle filter, followed by the Forward Filtering Backward Sampling (FFBS) procedure from \citet{godsillMonteCarloSmoothing2004}.
SMC works by approximating the distribution of the state $Z_n \equiv (X_{n}, V_{n}, S_n)$ conditional on all data up to and including index $n$, denoted as $Y^n$.
These approximations are formed by updating simulated particles targeting the distribution at index $n-1$ to particles targeting index $n$.
Starting with an equally-weighted sample $\tilde Z_{i, {n-1}}, i = 1, \dots, K$ targeting $\mP_\theta(Z_{{n-1}}| Y^{n-1})$,
new particles at index ${n}$ are drawn from a proposal distribution $Z_{n} \sim q(\cdot | \tilde Z_{i, {n-1}})$.
These are then reweighted and resampled to target the filtered distribution $\mP_\theta(Z_{i, n}| Y^n)$.
With enough particles, SMC gives an arbitrarily close approximation to the filtered distribution at each $n$.
To recover the smoothed distribution of each state conditional on all data points, $Z_{1:N} | Y^N$, we employ backward sampling over the existing samples from SMC using the FFBS algorithm.

Choosing a good proposal distribution is important for efficient inference with SMC.
A simple choice is to propose particles from the transition equation $f_{\theta}$, which is blind to future observations \citep{gordonNovelApproachNonlinear1993}.
This so-called ``bootstrap filter'' works well in many settings but poses a problem for the Argo data,
because, as mentioned earlier, the GPS measurement noise in Equation \ref{eq:5} is much lower than the transition noise in Equation \ref{eq:3}.
The chance that a proposal from the transition distribution will land close to the actual observed point is quite low.
As a result, the importance weights on each particle will be very uneven, frequently placing all probability mass on a single particle.
This leads to sample degeneracy and high variance in both the estimates of the predicted path and the log-likelihood of the model for inference.

There are a few properties of ArgoSSM that can be utilized to form an efficient proposal distribution.
First, the ice-avoidance algorithm state $S_n$ can be integrated out by calculating $P(S_{i, n} = k | X_{i, 1:n})$ recursively:
\begin{equation} \label{eq:ice_integration}
\begin{split}
    P(S_{n} = k | X_{1:n}, A_{1:n}) &= \frac{P(A_n | S_n = k) P(S_n = k |  X_{1:n}, A_{1:n-1}) }{\sum_ {k^\prime} P(A_n | S_n = k^\prime) P(S_n = k^\prime |  X_{1:n})} \\
    P(S_n = k |  X_{1:n}, A_{1:n-1}) &= \sum_{k^\prime} P(S_n = k | X_n, S_{n-1} = k^\prime) P(S_{n-1} = k^\prime | X_{1:n-1},  A_{1:n-1})
\end{split}
\end{equation}
Next, with $S_n$ integrated out, we formulate a proposal distribution to update the location $X_n$ and velocity $V_n$ given the previous states $X_{n-1}$ and $V_{n-1}$.
The ideal proposal distribution is equal to the posterior distribution conditional on all available data, denoted $Y^N$, and previous latent variables, as this will lead to importance sampling weights that are exactly even \citep{guarnieroIteratedAuxiliaryParticle2017}:
\begin{equation}
  \label{eq:ideal_proposal}
  q(X_{n}, V_{n} | X_{{n-1}}, V_{{n-1}}) = \mP(X_{n}, V_{n} | X_{{n-1}}, V_{{n-1}}, Y^N) .
\end{equation}
The right-hand side of Equation (\ref{eq:ideal_proposal}) is intractable, but it offers guidance: a good proposal will be as close as possible to the posterior that looks ahead at all future data points.
For ArgoSSM, we can use the tractable posterior distribution from the AR model as a proposal distribution in each step of SMC:
\begin{equation}\label{eq:proposal_ar}
  q(X_{n}, V_{n} | X_{n-1}, V_{n-1}) = \mP_{\theta, \text{AR}}(X_{n}, V_{n} | X_{n-1}, V_{n-1}, Y^N) .
\end{equation}
Here, $\mP_{\theta, {\text{AR}}}$ is the posterior distribution of the AR model.
The parameters of the AR model are set to be equal to the equivalent parameters of ArgoSSM.
The distribution $\mP_{\text{AR}}(X_{n}, V_{n} | X_{n-1}, V_{n-1},  Y_{A})$ is Gaussian,
so its mean and covariance can be efficiently calculated with a Kalman smoother \citep{lopesParticleFiltersBayesian2011}.
Finally, at each time-step, we incorporate PV into the proposal by multiplying the mean and covariance by the same matrix $B$ that appears in Equation \ref{eq:pv_posterior}.

\begin{figure}[t]
  \centering
(A)\includegraphics[width=0.6\linewidth]{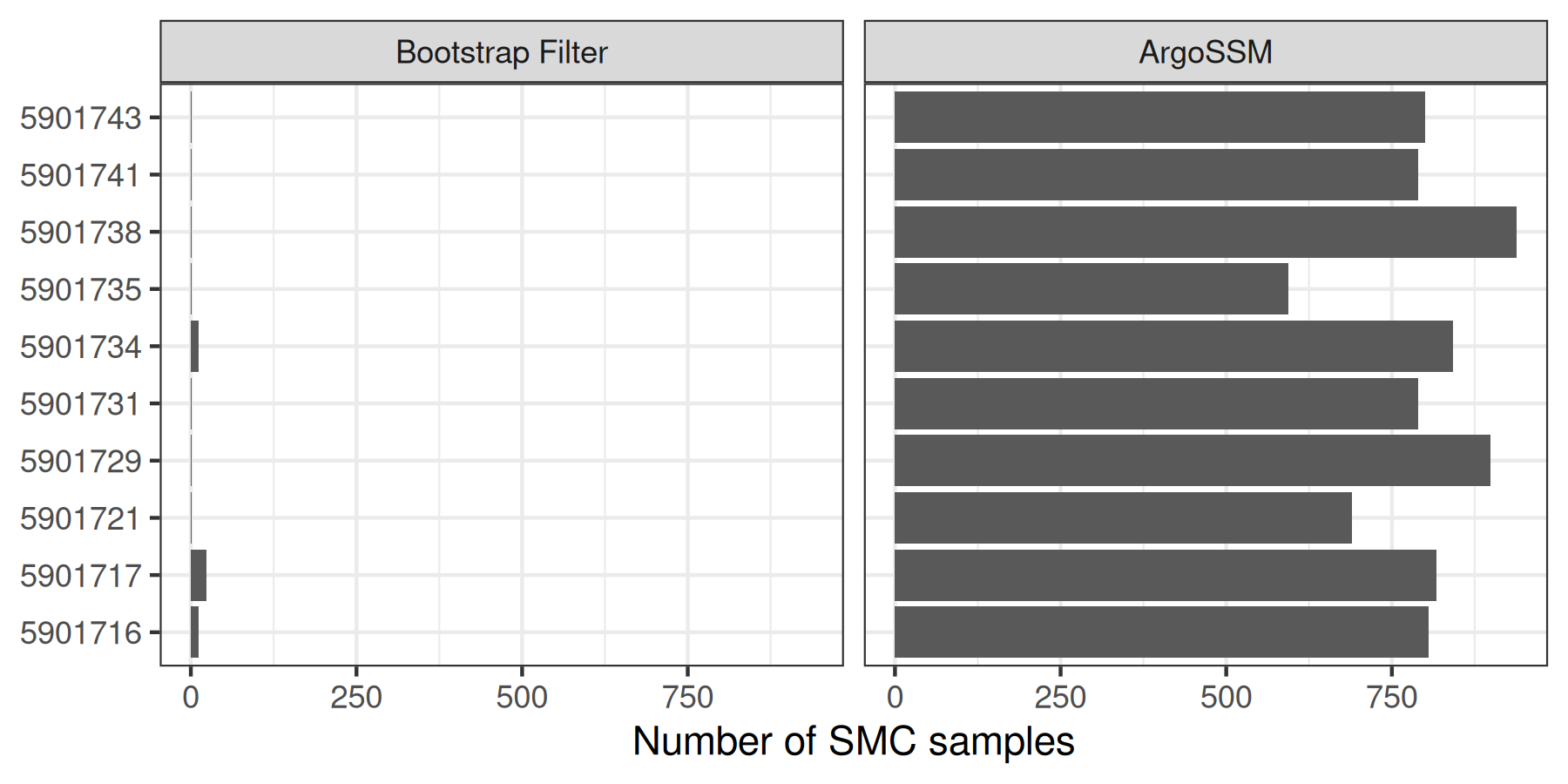}
(B)\includegraphics[width=0.3\linewidth]{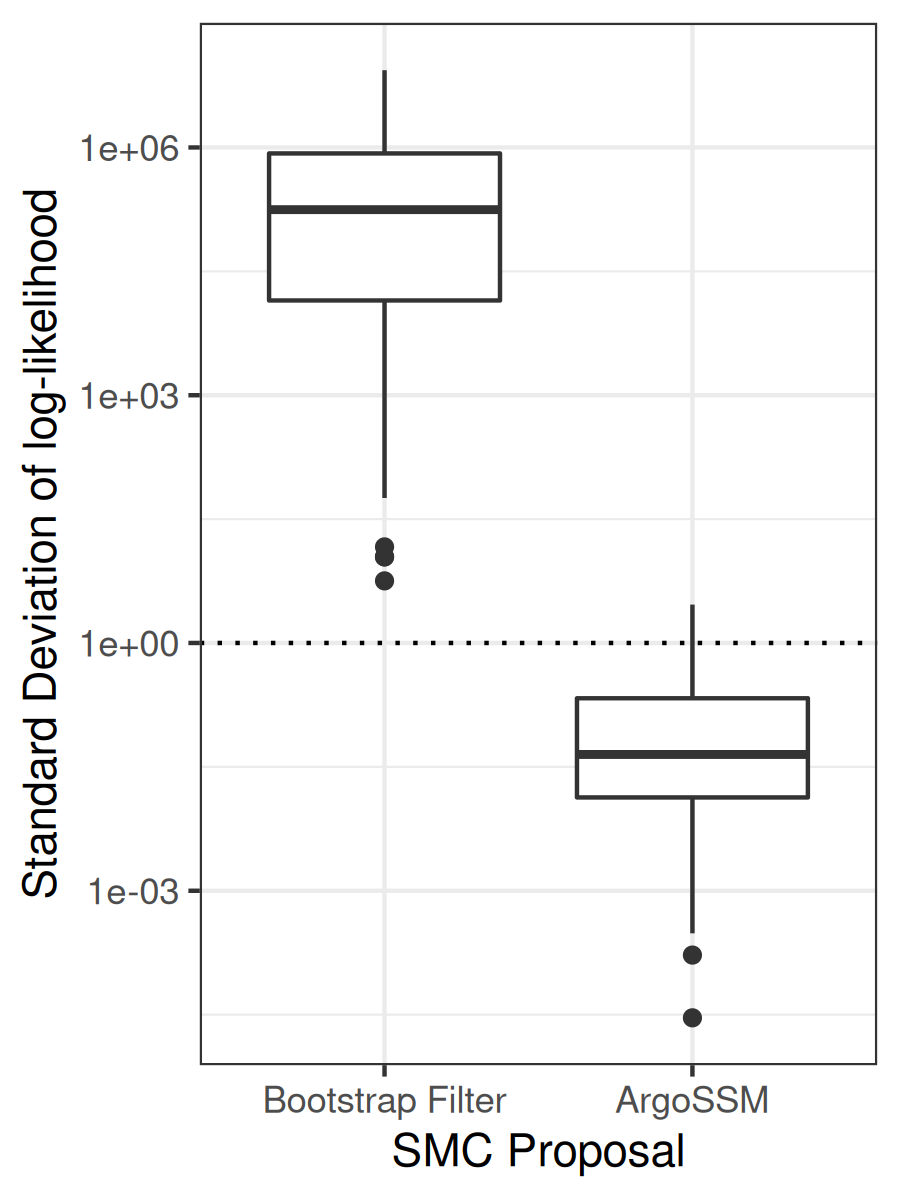}
  \caption{
    (A) Effective Sample Size (ESS) comparison from 1000 particles at the final observation for the bootstrap particle filter versus our adapted particle filter across each float used in the first holdout experiment. Floats are identified by their World Meteorological Organization (WMO) number. (B) The standard deviation of the estimated log-likelihood calculated across each float used in the first holdout experiment. The dotted line represents the target threshold for inference.}
  \label{fig:ess_comparison}
\end{figure}
We use this proposal within the auxiliary particle filter of \citet{pittFilteringSimulationAuxiliary1999} and likelihood estimate of \citet{pittSmoothParticleFilters2002}, following the helpful description given in Algorithm 1 of \citet{pittPropertiesMarkovChain2012}.
By looking ahead to all high signal-to-noise ratio (SNR) GPS measurements and directly incorporating local PV information, our proposal allows for a more efficient sampling procedure than the baseline bootstrap particle filter \citep{gordonNovelApproachNonlinear1993}.
Unlike other look-ahead methods such as \citet{linLookaheadStrategiesSequential2013}, our method takes advantage of the fact that the AR model is close to the ArgoSSM model and is fully tractable via the Kalman smoother.
We also utilize twisted target distributions at each index $n$ \citep{guarnieroIteratedAuxiliaryParticle2017, naessethElementsSequentialMonte2019a}, which amounts to weighting each particle by the ratio of predictive likelihoods from the AR model $\frac{\mP_{\theta, \text{AR}} (Y_{n:N} | X_{i, n-1}, V_{i, n-1})}{\mP_{\theta, \text{AR}} (Y_{n-1:N} | X_{i-1, n-1}, V_{i, n-1})}$ before resampling.
As a consequence of Proposition 2 from \citet{guarnieroIteratedAuxiliaryParticle2017}, estimating the AR model with this procedure yields an estimate of the likelihood.
This can also be verified through careful calculation of weights within SMC.
In particular, all of the stochastic terms cancel in the final estimated log-likelihood, leaving exactly the desired value $\mP_\theta(Y^N)$.

While the likelihood is no longer exact after incorporating the non-linearites from ice cover and PV conservation, the adapted SMC procedure still yields low-variance estimates of the model likelihood.
Figure \ref{fig:ess_comparison} shows an empirical comparison between the bootstrap filter and the proposal used in AR across various floats.
As expected, the bootstrap filter produces relatively few effective samples due to the high signal-to-noise ratio (SNR) in the data.
This translates into unacceptably high variability in the Monte Carlo estimates of the model likelihood.
With our proposal, the effective sample size is much higher, and the standard deviation in the estimated log-likelihood is well below 1 for the vast majority of floats, which is acceptable for parameter inference via SMC$^2$ described next.

\subsection{Bayesian parameter estimation via SMC$^{2}$}
\label{sec:param-estim-smc2}
\begin{table}[t]
\begin{center}
\begin{tabular}{r|l}
  Parameter & Prior Distribution \\
  \midrule
$\alpha$ & $\text{Beta}(8.9, 0.99)$ \\
$\sigma_{\text{PV}}^2$ & \text{LogNormal}(0.0, 3.0) \\
$v_{0, 1}$ & $\text{Normal}(0.0, 0.01)$ \\
$v_{0, 2}$ & $\text{Normal}(0.0, 0.01)$ \\
$p^{MAR}$ & $\text{Beta}(1.0, 9.0)$ \\
$p^{TPR}$ & $\text{Beta}(9.0, 1.0)$ \\
$p^{TNR}$ & $\text{Beta}(9.0, 1.0)$ \\
  \end{tabular}
\end{center}
    \caption{Table of priors used for each float parameter. The priors for the transition covariance $\Sigma_X$ and observation variance $\Sigma_Y$ were set by fitting a gamma distribution to maximum-likelihood estimates of the AR model on a small holdout set of floats.
    }
    \label{tbl:priors}
\end{table}

In addition to the uncertainty in the state variables, there is also uncertainty in the parameters which govern the dynamics, which we collectively refer to as $\theta$.
For ArgoSSM, these parameters are the initial parameters $\mu_{1}$ and $\Sigma_{1}$; the transition parameters $\Sigma_{X}$, $v_{0}$, $\alpha$, $\sigma^2_{PV}$, and $\Sigma_{V}$; the ice avoidance parameters $p^{\text{TPR}}, p^{\text{TNR}}, p^{\text{MAR}}$; and the observation variance $\Sigma_{Y}^2$.

Estimating parameters for a general state-space model is challenging,
because the model likelihood, $\mP(Y_{A} | \theta)$, requires evaluating an intractable integral over all state variables.
A key feature of SMC is that it provides an unbiased estimate of the likelihood, which can be used for either maximum likelihood estimation or Bayesian inference on $\theta$. See \citet{lopesParticleFiltersBayesian2011} for an overview of some of these approaches.

Our goal is to infer the posterior distribution of both $\theta$ and the state variables for each float.
To start, we equip each model parameter with a prior distribution as shown in Table \ref{tbl:priors}.
Then, to sample the posterior distribution of parameters on each float, we implement the SMC$^{2}$ procedure introduced in \cite{chopinSMC2EfficientAlgorithm2013}.
Starting with samples from the prior $\mP(\theta)$, SMC$^{2}$ sequentially targets intermediate distributions $\mP_{1}(\theta), \dots, \mP_{K}(\theta)$ via importance sampling such that $\mP_{K} (\theta)$ is the posterior distribution.
Each intermediate distribution is constructed such that the Kullback-Leibler (KL) distance between adjacent distributions is small.
To pick the sequence of distributions, we use the likelihood tempering scheme described in \cite{duanDensityTemperedMarginalizedSequential2015}.
Each distribution is defined by exponentiating the likelihood with a value $\xi$ between $0$ and $1$:
\begin{equation}
  \label{eq:7}
  \begin{split}
    \mP_{{k}}(\theta | Y_A) \propto \mP(\theta) \mP(Y_A | \theta)^{\xi_{k}}, \text{where} \\
    \xi_{0} = 0 < \xi_{1} < \cdots < \xi_{K} = 1 .
\end{split}
\end{equation}
When the importance weights drop below a set effective sample size (ESS), we resample and use particle Markov chain Monte Carlo (PMCMC) \citep{andrieuParticleMarkovChain2010} to propose new $\theta$ to rejuvenate the sample.

\section{Propagating uncertainty to spatiotemporal estimates}
\label{sec:temp-salin-field}

Through probabilistic modeling, ArgoSSM enables principled use of all data captured by Argo floats, even if the locations are missing.
In turn, ArgoSSM allows analysis in previously underexplored areas of the ocean.
For example, \cite{gray_global_2014} remove parts of the Weddell Gyre and other areas from their global estimates of ocean circulation due to lack of data, and both \cite{reeve_gridded_2016} and \cite{reeve_horizontal_2019} found it difficult to resolve seasonal effects in their models due to winter ice cover blocking many of the GPS measurements.

To demonstrate the benefits of ArgoSSM in downstream analysis of ocean properties, we consider two spatiotemporal estimation tasks: (1) temperature and salinity from Argo float data and (2) circulation estimation from Argo trajectory data.
In each of these tasks, the goal is to use the irregularly-spaced Argo data to predict at unobserved locations, often on a regular grid.
The temperature and salinity estimation problem is scientifically relevant and well-studied; see \cite{roemmich_20042008_2009} for a standard approach in oceanography and \cite{reeve_gridded_2016} for analysis specific to the Weddell Gyre.
Approaches for this problem roughly correspond to the methodology for the task of spatial prediction in spatial statistics \citep{cressie_statistics_1993}, with specific application to Argo demonstrated in \cite{kuusela_locally_2018}.
For circulation (i.e., velocity) estimation, the Argo float trajectories between consecutive profiles are used to form an estimate of the horizontal ocean circulation at the parking depth (about 1 kilometer under the ocean surface) of the floats. 
Since the trajectory data is the primary object of interest here, missing location data can be more detrimental, and previous analyses choose to mask or discard missing trajectory data \citep{gray_global_2014, reeve_horizontal_2019}.
However, with ArgoSSM, velocity estimates are available at all time points and can be directly used since they are state variables.

We describe our model for temperature, though we use similar approaches for salinity and ocean circulation.
We combine data from all floats, using $j = 1, \dots, J$ to index the float. At indices $n = 1, \dots, N_j$, and conditional on the locations $X = \{X_{j, n}\}$ taken at times $\{t_{j,n}\}$, the profile temperatures $T_{j,n}$ at a particular depth in the ocean are generated from the following spatial distribution:
\begin{equation}
  \label{eq:8}
  T_{n} = \mu(X_{j, n}, t_{j,n}) + \kappa(X_{j, {n}}, t_{j,n}) + \epsilon_{T, j, n},
\end{equation}
where $\mu$ is a smooth, deterministic mean function; $\kappa$ is a zero-mean, spatiotemporally correlated random field; and $\epsilon_{T,j,n}$ is measurement error that follows a normal distribution $\mathcal{N}(0, \sigma^2_{\epsilon_T})$.
We estimate $\mu(\cdot, \cdot)$ using locally-constant regression with four seasonal dynamic functions to capture seasonality in the estimates and a bandwidth of $250$ kilometers.
Following this, the covariance structure of the residuals are analyzed assuming a multivariate Gaussian distribution and using a Mat\'ern covariance function in space and time \citep{guinness_isotropic_2016, kuusela_locally_2018}.
Parameter estimates of the covariance function are obtained using maximum likelihood estimation using a Vecchia's approximation which eases computational burdens for spatial modeling with a large number of observations \citep{GPvecchia,katzfuss_general_2021}.
For velocity estimation, slight adjustments are made to this procedure since fewer floats will be used: a locally-constant estimator with no time dynamics was used for the mean estimation with a bandwidth of $400$ kilometers, and the use of time in the covariance structure was removed so that we estimate one time-averaged value at each location.

With missing locations, estimation is done by first drawing the locations $X$ from the posterior distribution of ArgoSSM, estimating parameters governing $\mu$, $\kappa$, and $\epsilon$ conditional on each location set, and then predicting from the distribution of temperature at an unobserved location conditional on the observed data.
By repeating this process for multiple samples of $X$ from ArgoSSM, one can capture the influence of the location uncertainty into the downstream predictions of temperature in a spirit similar to multiple imputation \citep[see][]{rubin_multiple_1987}.
We make two simplifying assumptions about the posterior distribution of floats from ArgoSSM: the temperatures ($T$) are weakly informative about the locations, and the dependence between floats is accounted for in the available data:
\begin{equation}
  \label{eq:4}
  \mP (\{X_{k, 1:N_{k}}, k = 1, \dots, K\}| Y^{N_k}, T) = \prod_{k = 1}^{K} \mP(X_{k, 1:N_{k}} | Y)
\end{equation}
In reality, the trajectories of floats in similar spatial areas are positively correlated, which means the assumptions in \autoref{eq:4} will lead to an over-dispersed joint distribution of locations, which is acceptable for the task at hand.
Incorporating dependencies between floats is a potential avenue for future work.

For spatiotemporal estimation tasks, ArgoSSM offers two key advantages over standard interpolation approaches for recovering missing locations.
First, as demonstrated in heldout data experiments in Section \ref{sec:experiments}, ArgoSSM more accurately predicts float trajectories.
Second, ArgoSSM propogates uncertainty in location data into downstream estimates, capturing a source of variability which would otherwise be ignored.
After completing this analysis for multiple sample outputs from ArgoSSM, we can form estimates of the expectation $E(T_{j,n}) = E(E(T_{j,n} | X))$ (using the law of total expectation) that takes into account the uncertainty in the missing locations.
More critically, as illustrated by the total law of variation, ArgoSSM allows us to quantify the full uncertainty in downstream estimates:
\begin{equation}
  \mathrm{Var}(T_{j,n}) = E(\mathrm{Var}(T_{j,n} | X)) + \mathrm{Var}(E(T_{j,n} | X))\label{eq:9} .
\end{equation}
With fixed or imputed locations, only the first component of the total uncertainty would be accounted for in final estimates, and it would not be properly averaged over the distribution of $X$.
With ArgoSSM, the second component is accounted for as well.
As we will see in Section \ref{sec:spatiotemporal-inference}, these two components of uncertainty have quite different spatial characteristics.
Areas with a low conditional variance can nonetheless have a high variance in their estimates because most floats in that area were under ice.

\section{Analysis of Argo floats in the Southern Ocean}

\label{sec:experiments}
As a case study for our method, we consider $46$ floats that traversed a specific area in the Southern Ocean, the Weddell Gyre, from 2002 to 2020
\footnote{The float temperature, salinity, and trajectory data were downloaded from the Argo database on July 11, 2020}.
In this setting, our task is two-fold.
First, we aim to reconstruct accurate trajectories of where floats went while under ice.
Because it is impossible to verify how well ArgoSSM does on missing data points, we evaluate its performance on heldout data points adjacent to under-ice periods (Section \ref{sec:holdout-experiments}).
Second, we illustrate how ArgoSSM is useful for spatiotemporal estimation tasks by uncovering an additional source of uncertainty (Section \ref{sec:spatiotemporal-inference}).
This applies both to measurements of temperature and salinity taken by the sensors and velocity estimates $V_n$ calculated as part of the probabilistic model.
Throughout this analysis, we draw $200$ samples of model parameters $\theta$ and use $K=2500$ particles per parameter sample to estimate the posterior distribution of float trajectories.
The code used for performing this analysis is publicly available on Github at \href{https://github.com/dereklhansen/ArgoSSM}{https://github.com/dereklhansen/ArgoSSM}.

\subsection{Holdout experiment} \label{sec:holdout-experiments}

To evaluate the empirical predictive performance of ArgoSSM under ice, we observe how well the model predicts heldout GPS points.
A single observation is heldout immediately before or after the float has been under ice cover for a minimum of 36 days.
This led to 197 independent holdout trials from 44 of the 48 floats.
On each holdout trial, we compare ArgoSSM to several baseline models.
These baseline models include the simple random walk (RW) model
defined in Equation \ref{eq:1} and the autoregressive (AR) model defined in Equations \ref{eq:2} and \ref{eq:3}.
To further evaluate which parts of ArgoSSM (AR+Ice+PV) lead to improvement, we consider a variant that just includes the ice-cover data (AR+Ice).
Finally, we compare to the PV interpolation method from \citet{chamberlain_observing_2018}.

The results of these two holdout experiments are summarized in Table \ref{tab:rmse}.
Across each holdout, we compare each model's prediction to the known true GPS measurements and calculate the root mean squared error (RSME) in kilometers.
We also consider the median squared error, which suppresses the impact of extreme values.
The AR and ArgoSSM outperform the baseline models of linear interpolation and PV interpolation in both measures.
While incorporating ice-cover and PV into the model led to a slightly higher median RMSE than the baseline AR model, mean RMSE is reduced by a much larger amount, suggesting that ArgoSSM is particularly useful in reducing holdouts with larger sources of error.
\begin{table}[]
\begin{center}
     \begin{tabular}{rcc}
    Model & RMSE & Median \\
    \midrule
    PV Interpolation \citep{chamberlain_observing_2018} & 33.1 & 13.2 \\
    Linear Interpolation (RW) & 18.0 & 11.4 \\
    \midrule
    AR & 16.1 & \textbf{8.58} \\
    AR+Ice & 15.7 & 8.92 \\
    ArgoSSM (AR+Ice+PV) & \textbf{15.4} & 8.64 \\
  \end{tabular}
\end{center}
\caption{
The observed prediction error of each method in kilometers in heldout data experiments. In the second, the floats had gone under the ice and additional measurements were held out. Note that the RW (``Random Walk'') predictions are equivalent to linear interpolation.
}
\label{tab:rmse}
\end{table}
\begin{figure}[]
  \centering
\includegraphics[width=\linewidth]{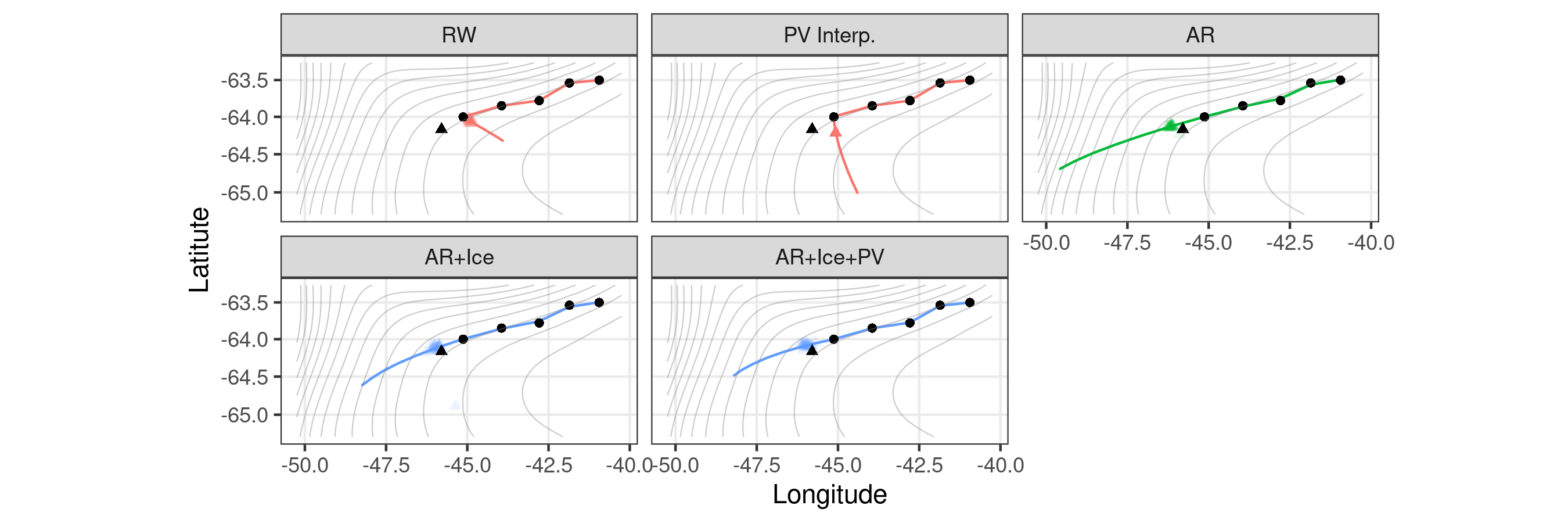}
  \caption{Comparison of each different model on Argo float 5901717 in predicting a heldout GPS measurement.
  The colored lines are the predicted paths from each model. Each black circle is a GPS observation given to the model and the black triangle is the heldout point.
  }
  \label{fig:chamberlain_posterior}
\end{figure}

Figure \ref{fig:chamberlain_posterior} provides a visual comparison of the performance of each model on float $5901717$
\footnote{With this float and throughout, we refer to floats through their World Meteorological Organization (WMO) number.},
which was also investigated in \citet{chamberlain_observing_2018}.
The random walk (RW) model, which amount to linear interpolation, performs poorly
because it uses no local information about momentum, ice-cover, nor potential vorticity (PV).
However, despite utilizing potential vorticity (PV), the PV interpolation method from \cite{chamberlain_observing_2018} also misses the target.
We see the autoregressive model (AR) does better than the baselines because it takes into account momentum.
Compared to the difference between AR and the baseline models,
the overall impact to predictive performance of adding ice cover and PV information is modest, but it makes a noticeable difference in Figure \ref{fig:chamberlain_posterior}.
Here, the AR model predicts the float swinging out too far.
Including either or both ice-cover and PV information corrects for this, leading to a more accurate prediction.

\subsection{Inference of spatiotemporal properties}\label{sec:spatiotemporal-inference}
With the posterior distribution of float trajectories, we show how the uncertainty in float locations shown in Figure \ref{fig:all_floats} leads to additional uncertainty in estimated temperature, salinity, and velocity (circulation) fields that would otherwise be ignored by imputation methods.
As a challenging example for temperature and salinity, we calculate estimates on August 1, 2015, a date in the middle of winter where many profiles have missing locations.

First, as described in Section \ref{sec:param-estim-smc2}, we infer the posterior distribution of model parameters for each float separately.
Figure \ref{fig:parameter_posteriors} shows the marginal posterior distribution of parameters for each float.
This shows that the estimated parameter values for $\alpha$ and $\sigma^2_{\text{PV}}$ can vary significantly between floats.
For example, the high value of $\sigma^2_{\text{PV}}$ for floats 5905995 and 5905381 indicates that the PV effect was not as strong as float 5901717.
This illustrates the benefit of allowing different parameters per-float.
Figure \ref{fig:parameter_posteriors} also shows the inferred true-positive-rate (TPR) and true-negative-rate (TNR) ice detections.
Like for the other parameters, these rates can differ significantly from float to float.
The majority of floats display a higher TPR than TNR, highlighting the conservative nature of the ice-detection algorithm \citep{klatt_profiling_2007}.

Across multiple draws of model parameters, float locations are sampled using the FFBS method from \citet{godsillMonteCarloSmoothing2004}.
Figure \ref{fig:all_floats} (A) shows that the estimated locations have varying degrees of uncertainty depending on location, time since the last observation, and float-specific parameters.
The regional differences between locations are particularly striking.
In the southwestern part of the map, profiles are both more sparse and have more missing locations, which adds significant uncertainty to downstream estimates.
In Figure \ref{fig:all_floats} (B), we display the uncertainty in kilometers as a random variable over time for several floats.
We see that this uncertainty differs greatly between floats.
This also underscores the need to incorporate downstream measurements, as
median location uncertainty is as high as 200km for some floats.

\begin{figure}[p]
  \centering
\includegraphics[width=0.8\linewidth]  {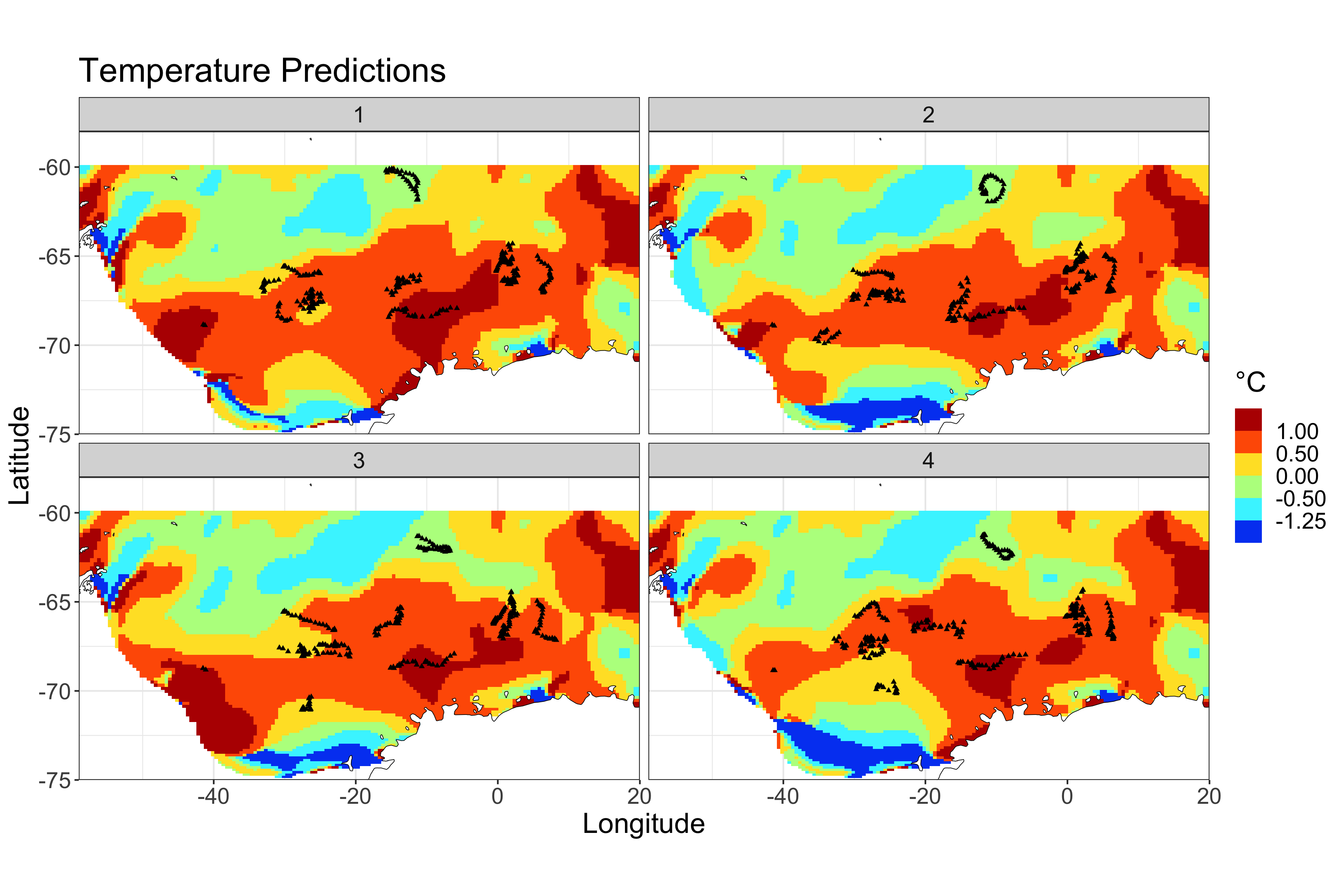}
\caption{Mean temperature estimates on August 1, 2015, taken on four different samples of locations. The black dots show the ArgoSSM sample of missing locations that was used to construct the estimate. Estimates were constructed for a pressure of 150 decibars, corresponding to about 150 meters deep in the ocean.}
  \label{fig:samples_temp}
\end{figure}
\begin{figure}[p]
  \centering
\begin{minipage}{0.45\linewidth}
  \includegraphics[width=0.8\linewidth]{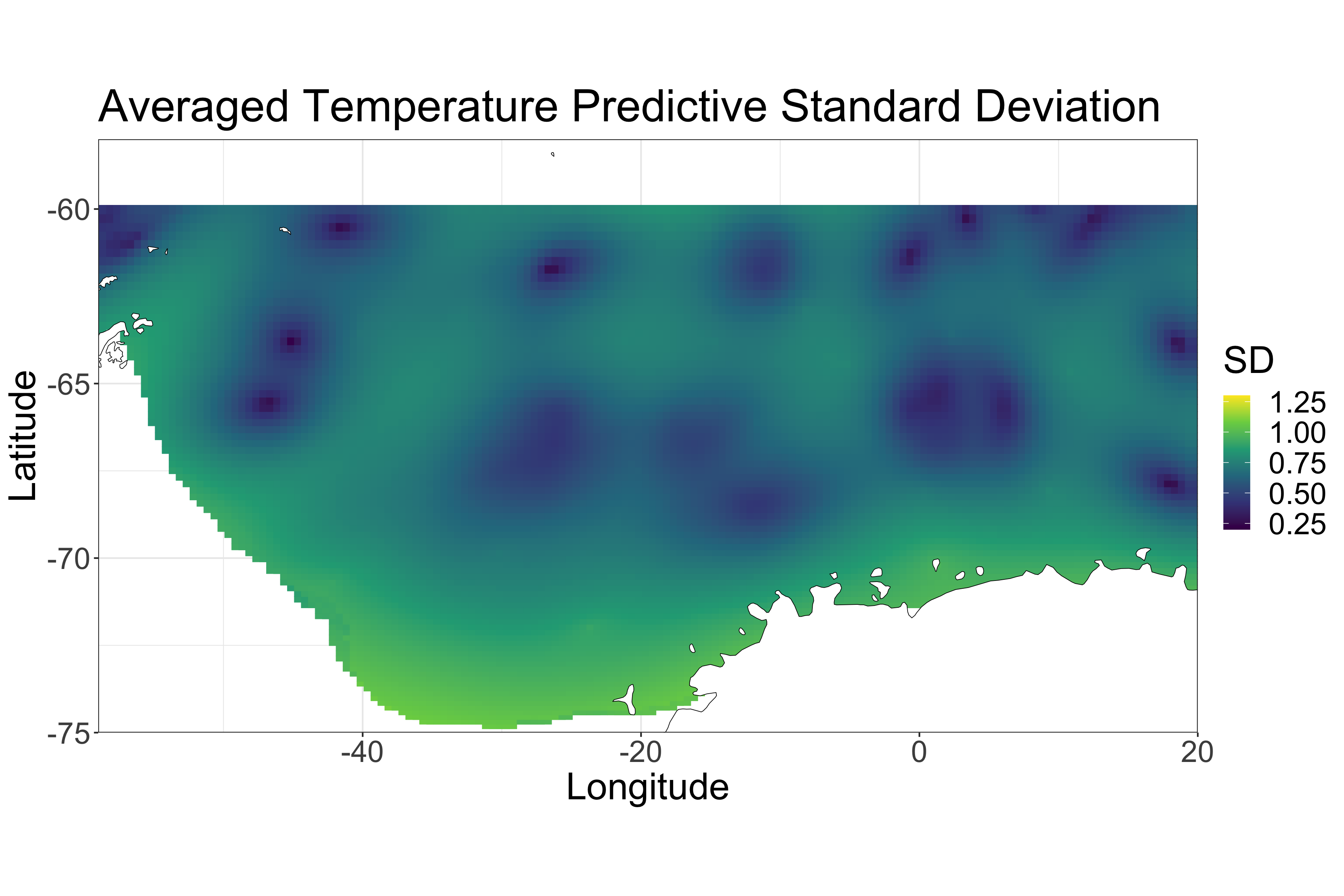}

  (A)
\end{minipage}
\begin{minipage}{0.45\linewidth}
  \includegraphics[width=0.8\linewidth]  {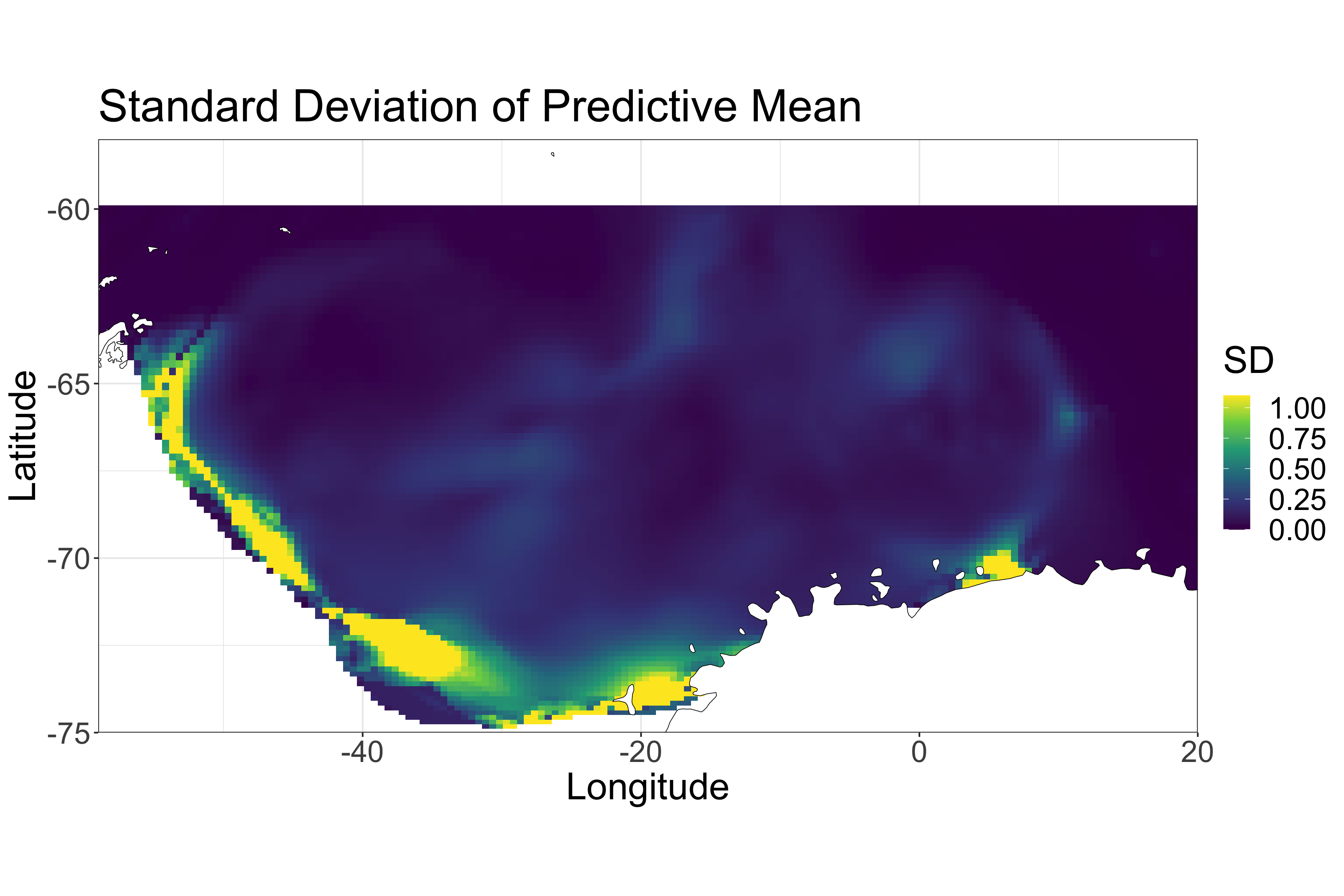}

  (B)
\end{minipage}

\begin{minipage}{0.45\linewidth}
  \includegraphics[width=0.8\linewidth]  {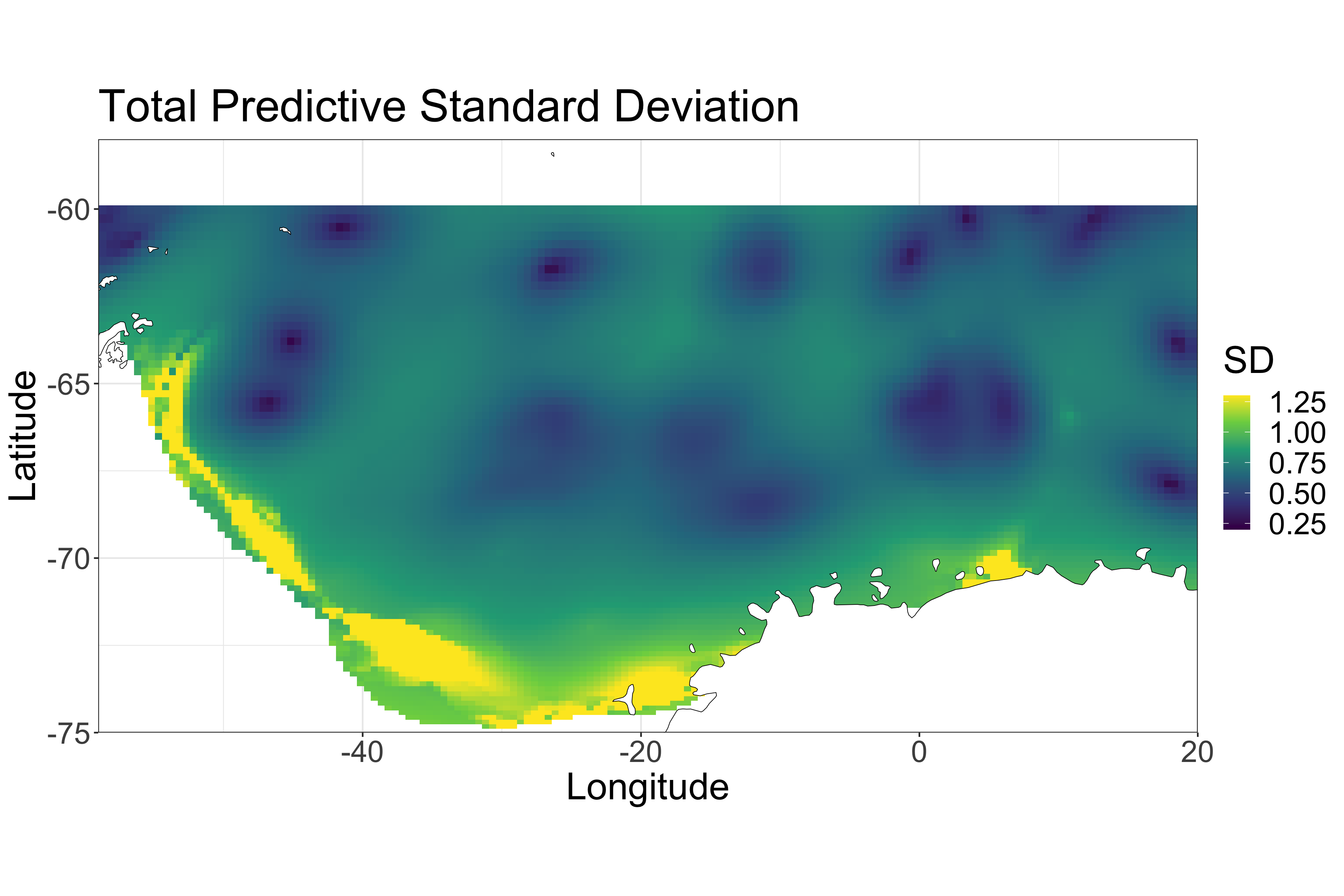}

  (C)
\end{minipage}

  \caption{Spatial uncertainty in temperature estimated on August 1, 2015. (A) Average standard deviation conditional on locations. (B) The standard deviation of mean temperature estimates between samples of locations. (C) Total standard deviation accounting for both (A) and (B).}
  \label{fig:spatial_uncertainty}
\end{figure}
\paragraph*{Temperature and salinity} For temperature and salinity, we focus on Argo data taken at 150 decibars of pressure, or about 150 meters deep in the ocean.
As described in Section \ref{sec:temp-salin-field}, we repeatedly estimate the temperature field conditional on samples of locations from the posterior distribution of ArgoSSM.
Figure \ref{fig:samples_temp} shows that estimated temperatures do not change much between samples in areas that have many observations.
However, in certain regions, there are noticeable differences between samples that come solely from the uncertainty in locations.
Figure \ref{fig:spatial_uncertainty} shows the breakdown in temperature uncertainty using 20 samples from ArgoSSM.
With one sample of locations, the conditional standard deviation of predicted temperatures depends on the estimated covariances and the proximity of nearby points.
Figure~\ref{fig:spatial_uncertainty}~(A) shows an estimate of the mean conditional standard deviation across the 20 location samples, which estimates the first quantity in Equation \ref{eq:9}.
In contrast, Figure \ref{fig:spatial_uncertainty} (B) shows the standard deviation of temperature estimates between samples, corresponding to the second quantity in Equation \ref{eq:9}.
Most areas in this plot have relatively low standard deviation compared to those in Figure \ref{fig:spatial_uncertainty} (A). However, there is a striking amount of uncertainty in the southern edges of the map, where there are few observations.
This uncertainty is large enough to significantly contribute to the total standard deviation shown in Figure~\ref{fig:spatial_uncertainty}~(C).
Without ArgoSSM, this contribution would be ignored.

Returning briefly to Figure~\ref{fig:spatial_uncertainty}~(A),
we also see that the areas with less predictive variability (near observed float locations) are more muted and indistinct when there is more location uncertainty for nearby floats.
That is, this component of the variance results from  properly averaging the spatially-predicted uncertainty across the distribution of missing locations.
As a result, using one fixed set of locations for Figure~\ref{fig:spatial_uncertainty}~(A) would lead to overconfidence near the fixed locations and underconfidence near other plausible float locations.
With ArgoSSM, improved estimates of Figure~\ref{fig:spatial_uncertainty}~(A) and Figure~\ref{fig:spatial_uncertainty}~(B) comprehensively characterize the variability of the estimated temperature field.

We plot the estimated salinity fields under four different samples from ArgoSSM in Figure \ref{fig:samples_psal}.
As with the temperature estimates, the location uncertainty has some effect on the estimated salinity field, especially near missing locations.
In Figure~\ref{fig:est_psal}~(A) and Figure~\ref{fig:est_psal}~(B), we plot the respective plots of the variability in salinity suggested by Equation \ref{eq:9}.
In general, we see similar results:  Figure~\ref{fig:est_psal}~(A) properly averages the prediction error over the distribution of missing locations, smoothed over areas with uncertainty in float locations, and Figure~\ref{fig:est_psal}~(B) represents uncertainty added based on the uncertainty in locations, which primarily contributes when there are few floats in the bottom-left of the plot.
Together, these describe the variance of salinity in a principled manner that incorporates our understanding of the missing locations through ArgoSSM.

\paragraph*{Ocean circulation} 

For estimates of ocean circulation, we only use floats with parking depths between 950 and 1050 meters to provide an estimate of the ocean circulation at approximately 1000 meters resulting in 26 floats, with locations plotted in Figure \ref{fig:veldata}.

\begin{figure}
\centering
\begin{minipage}{0.45\linewidth}
\includegraphics[width = 0.8\linewidth]{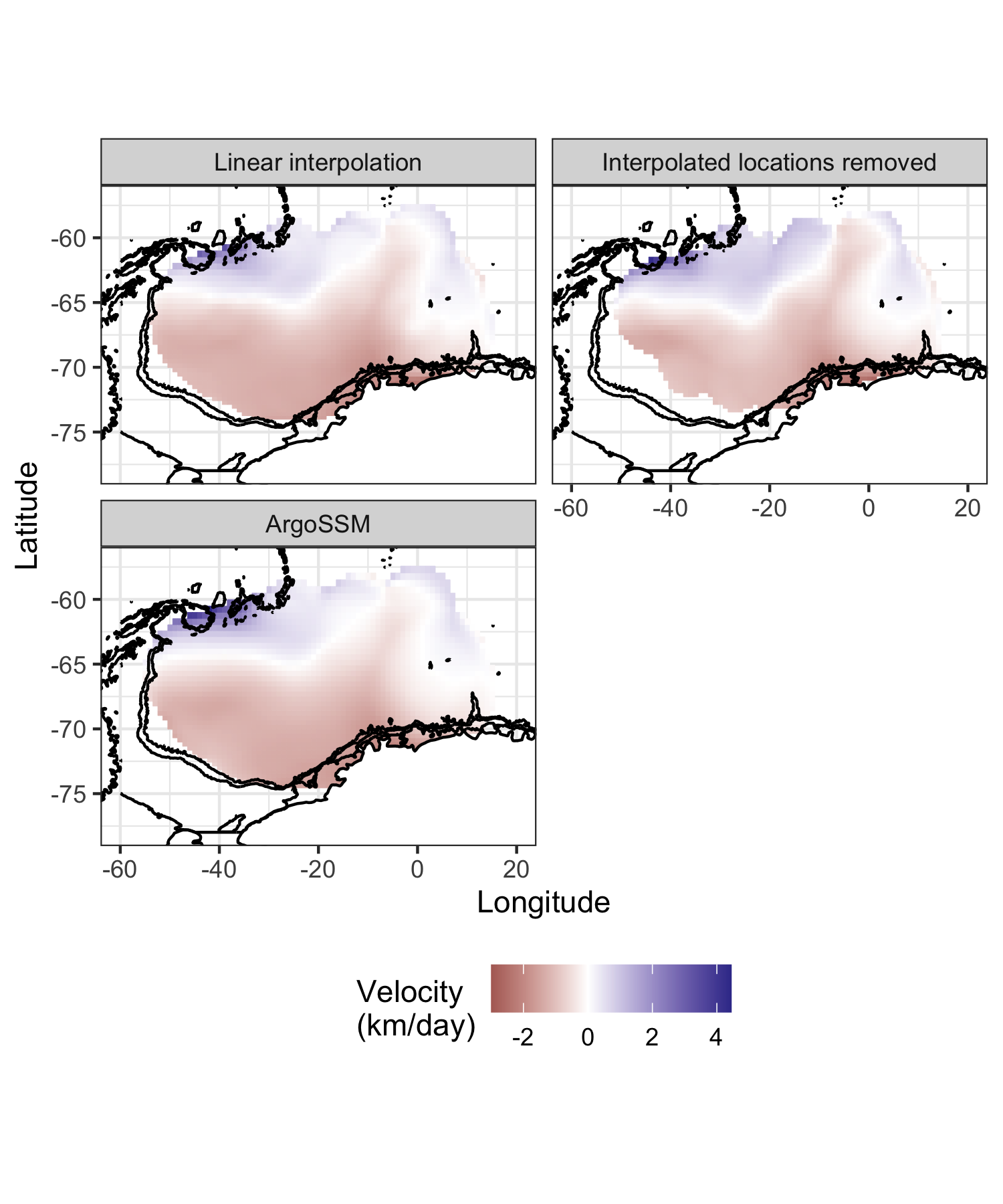}

  (A)
\end{minipage}
\begin{minipage}{0.45\linewidth}
\includegraphics[width = 0.8\linewidth]{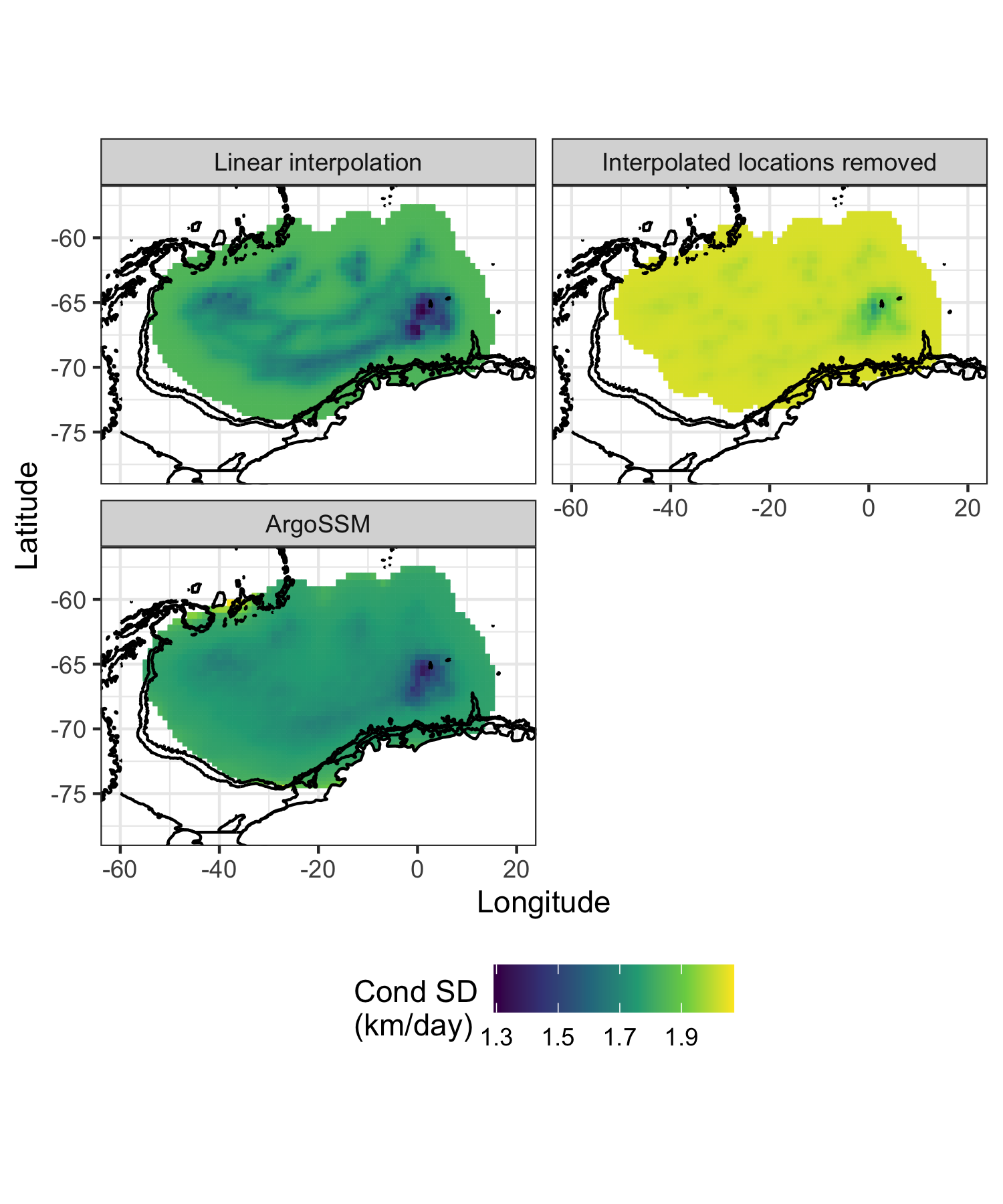}

  (B)
\end{minipage}
\caption{(A) Conditional expectation and (B) conditional standard deviation of zonal (east-west) velocity under different approaches. ``ArgoSSM'' refers to our estimates after using ArgoSSM to take into account the variability in the locations.}\label{fig:vel}
\end{figure}

Figure~\ref{fig:vel} shows results for the estimated zonal (east-west) velocity used to estimate ocean circulation.
As with temperature estimation, the zonal velocity is sensitive to missing locations.
In the left of Figure \ref{fig:vel}, we plot the conditional expectation based on the estimated spatial model for zonal velocity, where a negative value represents movement west while a positive value represents movement east.
We compare with linearly-interpolated locations, for which the float velocities are assumed constant while under-ice, as well as under-ice locations removed as done in \cite{reeve_horizontal_2019}.
While there are only slight differences between the methods in their mean estimates of velocity, the conditional variance estimates show more drastic variation (Figure \ref{fig:vel} B).
Simply removing the missing estimates leads to high uncertainty, while filling in the missing estimates via linear interpolation leads to overconfident final estimates.
Also, linear interpolation is more conservative on the basin's boundaries since Argo floats likely swing closer to the boundaries compared to the linearly interpolated paths.
Meanwhile, ArgoSSM provides a joint distribution of the positions and velocities for each float at times when profiles are collected, allowing the use of PV information while fully characterizing the uncertainty in the missing data.

\section{Discussion}\label{sec:discussion}
Before the Argo project, the Southern Ocean had been disproportionately under-observed compared to other areas of the world's oceans, especially during the winter  \citep{roemmich_20042008_2009}.
Improving the use of existing and future observations in this region through technological and methodological means at a low cost is an important scientific priority \citep{vernet_weddell_2019, riser_profiling_2018}.
A critical aspect of this is the use of profiles with missing location data while under ice cover.

ArgoSSM improves upon existing approaches for under-ice location estimation in two ways.
First, it more realistically models how floats drift, which is validated by heldout data experiments.
Second, through inferring a posterior distribution of locations, ArgoSSM accounts for uncertainty stemming from missing locations.
These improvements lend confidence to areas with low estimated uncertainty and signal areas where the variance is too high to draw conclusions.
This could focus future data-gathering endeavors to areas of maximum impact.

ArgoSSM will enable scientists to incorporate location or velocity uncertainty into their estimates.
ArgoSSM posterior samples can be used to evaluate the sensitivity to location uncertainty, which may vary from task to task \citep{riser_profiling_2018, chamberlain_observing_2018}.
Future work could use ArgoSSM estimates to develop an ``error-in-variables'' approach to specific estimation tasks, removing the need to repeatedly estimate under different samples of locations \citep{cervone_gaussian_2015}.
However, since this literature primarily focuses on independent and identically-distributed errors in the locations, such an approach may require care and methodological development.
In this setting, for the same float in the same winter, the missing locations can be strongly correlated.
We advocate for the sampling-based approach developed here since it considers this correlation and can be employed naturally by statisticians and scientists alike.

The current version of ArgoSSM may oversimplify the dynamics of float movement.
However, ArgoSSM is a flexible framework that can be improved with better physical models and expanded to accommodate more data as it becomes available.
For example, one could pair ArgoSSM with ocean circulation estimates instead of potential vorticity, though such estimates may be noisy in the winter due to sparse measurements \citep{reeve_horizontal_2019}.
Also, one could estimate model parameters and float positions using data from multiple floats simultaneously, which might further reduce the size of ArgoSSM's uncertainties.
Additional location measurements could be incorporated as they become available, such as RAFOS sound source data \citep{chamberlain_observing_2018}.
Through these iterative improvements, ArgoSSM can utilize all available information to best predict float locations, improving existing downstream tasks and enabling new ones.

\section*{Funding}
This material is based upon work supported by the National Science Foundation Graduate Research Fellowship Program under grants no. 1256260 and 1841052. Any opinions, findings, and conclusions or recommendations expressed in this material are those of the authors and do not necessarily reflect the views of the National Science Foundation.

\section*{Acknowledgments}
We would like to thank Stilian Stoev, Tailen Hsing, and Jeffrey Regier for helpful guidance and feedback on this work. We would also like to sincerely thank Sarah Gille and the Physical Oceanography group at Scripps Institution of Oceanography for suggesting this topic of research.

\section*{Data Availability Statement}
All data used in this work is publicly available. 

The Argo data is available online at: \url{https://argo.ucsd.edu/data/data-from-gdacs/}. These data were collected and made freely available by the International Argo Program and the national programs that contribute to it. (https://argo.ucsd.edu, https://www.ocean-ops.org). The Argo Program is part of the Global Ocean Observing System. 

The Southern Ocean State Estimate is available online at: \url{http://sose.ucsd.edu}. 
Computational resources for the SOSE were provided by NSF XSEDE resource grant OCE130007 and SOCCOM NSF award PLR-1425989.

Ice concentration data \citep{ice_data} is available at: \url{https://nsidc.org/data/seaice_index}.

\bibliographystyle{unsrtnat}

\begin{thebibliography}{39}
\providecommand{\natexlab}[1]{#1}
\providecommand{\url}[1]{\texttt{#1}}
\expandafter\ifx\csname urlstyle\endcsname\relax
  \providecommand{\doi}[1]{doi: #1}\else
  \providecommand{\doi}{doi: \begingroup \urlstyle{rm}\Url}\fi

\bibitem[Lyman and Johnson(2014)]{lyman_estimating_2014}
John~M. Lyman and Gregory~C. Johnson.
\newblock Estimating global ocean heat content changes in the upper 1800 m
  since 1950 and the influence of climatology choice.
\newblock \emph{Journal of Climate}, 27\penalty0 (5):\penalty0 1945--1957,
  2014.
\newblock ISSN 0894-8755, 1520-0442.
\newblock \doi{10.1175/JCLI-D-12-00752.1}.
\newblock URL \url{http://journals.ametsoc.org/doi/10.1175/JCLI-D-12-00752.1}.

\bibitem[Chang et~al.(2009)Chang, Rosati, Zhang, and
  Harrison]{chang_objective_2009}
You-Soon Chang, Anthony~J. Rosati, Shaoqing Zhang, and Matthew~J. Harrison.
\newblock Objective analysis of monthly temperature and salinity for the world
  ocean in the 21st century: {Comparison} with {World} {Ocean} {Atlas} and
  application to assimilation validation.
\newblock \emph{Journal of Geophysical Research}, 114\penalty0 (C2):\penalty0
  C02014, 2009.
\newblock ISSN 0148-0227.
\newblock \doi{10.1029/2008JC004970}.
\newblock URL \url{http://doi.wiley.com/10.1029/2008JC004970}.

\bibitem[Hosoda et~al.(2009)Hosoda, Suga, Shikama, and
  Mizuno]{hosodaGlobalSurfaceLayer2009}
Shigeki Hosoda, Toshio Suga, Nobuyuki Shikama, and Keisuke Mizuno.
\newblock Global surface layer salinity change detected by {{Argo}} and its
  implication for hydrological cycle intensification.
\newblock \emph{Journal of Oceanography}, 65\penalty0 (4):\penalty0 579--586,
  2009.
\newblock ISSN 1573-868X.
\newblock \doi{10.1007/s10872-009-0049-1}.

\bibitem[Argo(2020)]{argo2020}
Argo.
\newblock Argo float data and metadata from {Global} {Data} {Assembly} {Centre}
  ({Argo} {GDAC}), 2020.
\newblock URL \url{https://www.seanoe.org/data/00311/42182/}.

\bibitem[Klatt et~al.(2007)Klatt, Boebel, and Fahrbach]{klatt_profiling_2007}
Olaf Klatt, Olaf Boebel, and Eberhard Fahrbach.
\newblock A {profiling} {float}’s {sense} of {ice}.
\newblock \emph{Journal of Atmospheric and Oceanic Technology}, 24\penalty0
  (7):\penalty0 1301--1308, 2007.
\newblock ISSN 0739-0572, 1520-0426.
\newblock \doi{10.1175/JTECH2026.1}.
\newblock URL \url{http://journals.ametsoc.org/doi/abs/10.1175/JTECH2026.1}.

\bibitem[Chamberlain et~al.(2018)Chamberlain, Talley, Mazloff, Riser, Speer,
  Gray, and Schwartzman]{chamberlain_observing_2018}
Paul~M. Chamberlain, Lynne~D. Talley, Matthew~R. Mazloff, Stephen~C. Riser,
  Kevin Speer, Alison~R. Gray, and Armin Schwartzman.
\newblock Observing the {ice}-{covered} {Weddell} {Gyre} {with} {profiling}
  {floats}: {position} {uncertainties} and {correlation} {statistics}.
\newblock \emph{Journal of Geophysical Research: Oceans}, 123\penalty0
  (11):\penalty0 8383--8410, 2018.
\newblock ISSN 2169-9291.
\newblock \doi{10.1029/2017JC012990}.
\newblock URL
  \url{https://agupubs.onlinelibrary.wiley.com/doi/abs/10.1029/2017JC012990}.

\bibitem[Reeve et~al.(2019)Reeve, Boebel, Strass, Kanzow, and
  Gerdes]{reeve_horizontal_2019}
Krissy~Anne Reeve, Olaf Boebel, Volker Strass, Torsten Kanzow, and Rüdiger
  Gerdes.
\newblock Horizontal circulation and volume transports in the {Weddell} {Gyre}
  derived from {Argo} float data.
\newblock \emph{Progress in Oceanography}, 175:\penalty0 263--283, 2019.
\newblock ISSN 0079-6611.
\newblock \doi{10.1016/j.pocean.2019.04.006}.
\newblock URL
  \url{http://www.sciencedirect.com/science/article/pii/S0079661117302756}.

\bibitem[Gray and Riser(2014)]{gray_global_2014}
Alison~R. Gray and Stephen~C. Riser.
\newblock A global analysis of {Sverdrup} balance using absolute geostrophic
  velocities from {Argo}.
\newblock \emph{Journal of Physical Oceanography}, 44\penalty0 (4):\penalty0
  1213--1229, 2014.
\newblock ISSN 0022-3670, 1520-0485.
\newblock \doi{10.1175/JPO-D-12-0206.1}.
\newblock URL \url{http://journals.ametsoc.org/doi/10.1175/JPO-D-12-0206.1}.

\bibitem[Reeve et~al.(2016)Reeve, Boebel, Kanzow, Strass, Rohardt, and
  Fahrbach]{reeve_gridded_2016}
Krissy Reeve, Olaf Boebel, T.~Kanzow, V.~Strass, G.~Rohardt, and E.~Fahrbach.
\newblock A gridded data set of upper-ocean hydrographic properties in the
  {Weddell} {Gyre} obtained by objective mapping of {Argo} float measurements.
\newblock \emph{Earth System Science Data}, 8:\penalty0 15--40, 2016.
\newblock \doi{10.5194/essd-8-15-2016}.

\bibitem[Gray et~al.(2018)Gray, Johnson, Bushinsky, Riser, Russell, Talley,
  Wanninkhof, Williams, and Sarmiento]{gray_autonomous_2018}
Alison~R. Gray, Kenneth~S. Johnson, Seth~M. Bushinsky, Stephen~C. Riser,
  Joellen~L. Russell, Lynne~D. Talley, Rik Wanninkhof, Nancy~L. Williams, and
  Jorge~L. Sarmiento.
\newblock Autonomous {biogeochemical} {floats} {detect} {significant} {carbon}
  {dioxide} {outgassing} in the {high}-{latitude} {Southern} {Ocean}.
\newblock \emph{Geophysical Research Letters}, 45\penalty0 (17):\penalty0
  9049--9057, 2018.
\newblock ISSN 1944-8007.
\newblock \doi{https://doi.org/10.1029/2018GL078013}.
\newblock URL
  \url{http://agupubs.onlinelibrary.wiley.com/doi/abs/10.1029/2018GL078013}.

\bibitem[Wong et~al.(2020)Wong, Wijffels, Riser, Pouliquen, Hosoda, Roemmich,
  Gilson, Johnson, Martini, Murphy, Scanderbeg, Bhaskar, Buck, Merceur, Carval,
  Maze, Cabanes, André, and Poffa]{wong_argo_2020}
Annie P.~S. Wong, Susan~E. Wijffels, Stephen~C. Riser, Sylvie Pouliquen,
  Shigeki Hosoda, Dean Roemmich, John Gilson, Gregory~C. Johnson, Kim Martini,
  David~J. Murphy, Megan Scanderbeg, T.~V. S.~Udaya Bhaskar, Justin J.~H. Buck,
  Frederic Merceur, Thierry Carval, Guillaume Maze, Cécile Cabanes, Xavier
  André, and Hyuk-Min Poffa, N. ...~Park.
\newblock Argo {data} 1999–2019: {Two} {million} {temperature}-{salinity}
  {profiles} and {subsurface} {velocity} {observations} {from} a {global}
  {array} of {profiling} {floats}.
\newblock \emph{Frontiers in Marine Science}, 7, 2020.
\newblock ISSN 2296-7745.
\newblock \doi{10.3389/fmars.2020.00700}.
\newblock URL
  \url{https://www.frontiersin.org/articles/10.3389/fmars.2020.00700/full}.

\bibitem[Talley et~al.(2011)Talley, Pickard, Emery, and
  Swift]{talley_chapter_2011}
Lynne~D. Talley, George~L. Pickard, William~J. Emery, and James~H. Swift.
\newblock Chapter 7 - {Dynamical} {processes} for {descriptive} {ocean}
  {circulation}.
\newblock In \emph{Descriptive {Physical} {Oceanography} ({Sixth} {Edition})},
  pages 187--221. Academic Press, Boston, 2011.
\newblock ISBN 978-0-7506-4552-2.
\newblock \doi{10.1016/B978-0-7506-4552-2.10007-1}.
\newblock URL
  \url{https://www.sciencedirect.com/science/article/pii/B9780750645522100071}.

\bibitem[Fetterer et~al.(2017)Fetterer, Knowles, Meier, Savoie, and
  Windnagel]{ice_data}
F.~Fetterer, K.~Knowles, W.~N. Meier, M.~Savoie, and A.~K. Windnagel.
\newblock \emph{Sea Ice Index: Sea Ice Concentration}, 2017.
\newblock Boulder, Colorado USA. NSIDC: National Snow and Ice Data Center.
  Version 3, updated daily.

\bibitem[Gelman et~al.(2015)Gelman, Carlin, Stern, Dunson, Vehtari, and
  Rubin]{gelmanBayesianDataAnalysis2015}
Andrew Gelman, John~B. Carlin, Hal~S. Stern, David~B. Dunson, Aki Vehtari, and
  Donald~B. Rubin.
\newblock \emph{Bayesian {{Data Analysis}}}.
\newblock {Chapman and Hall/CRC}, {New York}, third edition, July 2015.
\newblock ISBN 978-0-429-11307-9.
\newblock \doi{10.1201/b16018}.

\bibitem[Verdy and Mazloff(2017)]{verdy_data_2017}
A.~Verdy and M.~R. Mazloff.
\newblock A data assimilating model for estimating {Southern} {Ocean}
  biogeochemistry.
\newblock \emph{Journal of Geophysical Research: Oceans}, 122\penalty0
  (9):\penalty0 6968--6988, 2017.
\newblock ISSN 2169-9291.
\newblock \doi{https://doi.org/10.1002/2016JC012650}.
\newblock URL
  \url{http://agupubs.onlinelibrary.wiley.com/doi/abs/10.1002/2016JC012650}.

\bibitem[Fan and Gijbels(1996)]{fan_local_1996}
J.~Fan and I.~Gijbels.
\newblock \emph{Local {Polynomial} {Modelling} and its {Applications}},
  volume~66 of \emph{Monographs on {Statistics} and {Applied} {Probability}}.
\newblock Chapman \& Hall, London, 1996.
\newblock ISBN 0-412-98321-4.

\bibitem[Lopes and Tsay(2011)]{lopesParticleFiltersBayesian2011}
Hedibert~F. Lopes and Ruey~S. Tsay.
\newblock Particle filters and {{Bayesian}} inference in financial
  econometrics.
\newblock \emph{Journal of Forecasting}, 30\penalty0 (1):\penalty0 168--209,
  2011.
\newblock ISSN 02776693.
\newblock \doi{10.1002/for.1195}.

\bibitem[Godsill et~al.(2004)Godsill, Doucet, and
  West]{godsillMonteCarloSmoothing2004}
Simon~J Godsill, Arnaud Doucet, and Mike West.
\newblock Monte {Carlo} smoothing for nonlinear time series.
\newblock \emph{Journal of the American Statistical Association}, 99\penalty0
  (465):\penalty0 156--168, 2004.
\newblock ISSN 0162-1459, 1537-274X.
\newblock \doi{10.1198/016214504000000151}.

\bibitem[Gordon et~al.(1993)Gordon, Salmond, and
  Smith]{gordonNovelApproachNonlinear1993}
N.~J. Gordon, D.~J. Salmond, and A.~F.~M. Smith.
\newblock Novel approach to nonlinear/non-{{Gaussian Bayesian}} state
  estimation.
\newblock \emph{IEE Proceedings F (Radar and Signal Processing)}, 140\penalty0
  (2):\penalty0 107--113, 1993.
\newblock ISSN 2053-9045.
\newblock \doi{10.1049/ip-f-2.1993.0015}.

\bibitem[Guarniero et~al.(2017)Guarniero, Johansen, and
  Lee]{guarnieroIteratedAuxiliaryParticle2017}
Pieralberto Guarniero, Adam~M. Johansen, and Anthony Lee.
\newblock The iterated auxiliary particle filter.
\newblock \emph{Journal of the American Statistical Association}, 112\penalty0
  (520):\penalty0 1636--1647, 2017.
\newblock ISSN 0162-1459, 1537-274X.
\newblock \doi{10.1080/01621459.2016.1222291}.

\bibitem[Pitt and Shephard(1999)]{pittFilteringSimulationAuxiliary1999}
Michael~K. Pitt and Neil Shephard.
\newblock Filtering via {{Simulation}}: {{Auxiliary Particle Filters}}.
\newblock \emph{Journal of the American Statistical Association}, 94\penalty0
  (446):\penalty0 590--599, 1999.
\newblock ISSN 0162-1459, 1537-274X.
\newblock \doi{10.1080/01621459.1999.10474153}.

\bibitem[Pitt(2002)]{pittSmoothParticleFilters2002}
Michael~K. Pitt.
\newblock Smooth {{Particle Filters}} for {{Likelihood Evaluation}} and
  {{Maximisation}}.
\newblock The {{Warwick Economics Research Paper Series}} ({{TWERPS}}),
  {University of Warwick, Department of Economics}, 2002.

\bibitem[Pitt et~al.(2012)Pitt, Silva, Giordani, and
  Kohn]{pittPropertiesMarkovChain2012}
Michael~K. Pitt, Ralph dos~Santos Silva, Paolo Giordani, and Robert Kohn.
\newblock On some properties of {{Markov}} chain {{Monte Carlo}} simulation
  methods based on the particle filter.
\newblock \emph{Journal of Econometrics}, 171\penalty0 (2):\penalty0 134--151,
  2012.
\newblock ISSN 03044076.
\newblock \doi{10.1016/j.jeconom.2012.06.004}.

\bibitem[Lin et~al.(2013)Lin, Chen, and
  Liu]{linLookaheadStrategiesSequential2013}
Ming Lin, Rong Chen, and Jun~S. Liu.
\newblock Lookahead {{Strategies}} for {{Sequential Monte Carlo}}.
\newblock \emph{Statistical Science}, 28\penalty0 (1):\penalty0 69--94, 2013.
\newblock ISSN 0883-4237, 2168-8745.
\newblock \doi{10.1214/12-STS401}.

\bibitem[Naesseth et~al.(2019)Naesseth, Lindsten, and
  Sch{\"o}n]{naessethElementsSequentialMonte2019a}
Christian~A. Naesseth, Fredrik Lindsten, and Thomas~B. Sch{\"o}n.
\newblock Elements of {{Sequential Monte Carlo}}.
\newblock \emph{Foundations and Trends\textregistered{} in Machine Learning},
  12\penalty0 (3):\penalty0 307--392, November 2019.
\newblock ISSN 1935-8237, 1935-8245.
\newblock \doi{10.1561/2200000074}.

\bibitem[Chopin et~al.(2013)Chopin, Jacob, and
  Papaspiliopoulos]{chopinSMC2EfficientAlgorithm2013}
N.~Chopin, P.~E. Jacob, and O.~Papaspiliopoulos.
\newblock {{SMC$^2$}}: An efficient algorithm for sequential analysis of state
  space models.
\newblock \emph{Journal of the Royal Statistical Society: Series B (Statistical
  Methodology)}, 75\penalty0 (3):\penalty0 397--426, 2013.
\newblock ISSN 13697412.
\newblock \doi{10.1111/j.1467-9868.2012.01046.x}.

\bibitem[Duan and Fulop(2015)]{duanDensityTemperedMarginalizedSequential2015}
Jin-Chuan Duan and Andras Fulop.
\newblock Density-{{tempered marginalized Sequential Monte Carlo samplers}}.
\newblock \emph{Journal of Business \& Economic Statistics}, 33\penalty0
  (2):\penalty0 192--202, 2015.
\newblock ISSN 0735-0015, 1537-2707.
\newblock \doi{10.1080/07350015.2014.940081}.

\bibitem[Andrieu et~al.(2010)Andrieu, Doucet, and
  Holenstein]{andrieuParticleMarkovChain2010}
Christophe Andrieu, Arnaud Doucet, and Roman Holenstein.
\newblock Particle {{Markov}} chain {{Monte Carlo}} methods.
\newblock \emph{Journal of the Royal Statistical Society: Series B (Statistical
  Methodology)}, 72\penalty0 (3):\penalty0 269--342, 2010.
\newblock ISSN 13697412, 14679868.
\newblock \doi{10.1111/j.1467-9868.2009.00736.x}.

\bibitem[Roemmich and Gilson(2009)]{roemmich_20042008_2009}
D.~Roemmich and J.~Gilson.
\newblock The 2004–2008 mean and annual cycle of temperature, salinity, and
  steric height in the global ocean from the {Argo} {Program}.
\newblock \emph{Progress in Oceanography}, 82\penalty0 (2):\penalty0 81--100,
  2009.

\bibitem[Cressie(1993)]{cressie_statistics_1993}
Noel A.~C. Cressie.
\newblock \emph{Statistics for spatial data}.
\newblock Wiley series in probability and mathematical statistics. {Applied}
  probability and statistics. J. Wiley, New York, rev. ed. edition, 1993.
\newblock ISBN 978-0-471-00255-0.

\bibitem[Kuusela and Stein(2018)]{kuusela_locally_2018}
Mikael Kuusela and Michael~L. Stein.
\newblock Locally stationary spatio-temporal interpolation of {Argo} profiling
  float data.
\newblock \emph{Proceedings of the Royal Society A: Mathematical, Physical and
  Engineering Sciences}, 474\penalty0 (2220):\penalty0 20180400, 2018.
\newblock ISSN 1364-5021, 1471-2946.
\newblock \doi{10.1098/rspa.2018.0400}.
\newblock URL
  \url{https://royalsocietypublishing.org/doi/10.1098/rspa.2018.0400}.

\bibitem[Guinness and Fuentes(2016)]{guinness_isotropic_2016}
Joseph Guinness and Montserrat Fuentes.
\newblock Isotropic covariance functions on spheres: {Some} properties and
  modeling considerations.
\newblock \emph{Journal of Multivariate Analysis}, 143:\penalty0 143--152,
  2016.
\newblock ISSN 0047259X.
\newblock \doi{10.1016/j.jmva.2015.08.018}.
\newblock URL
  \url{https://linkinghub.elsevier.com/retrieve/pii/S0047259X15002109}.

\bibitem[Katzfuss et~al.(2020)Katzfuss, Jurek, Zilber, and Gong]{GPvecchia}
Matthias Katzfuss, Marcin Jurek, Daniel Zilber, and Wenlong Gong.
\newblock \emph{GPvecchia: scalable Gaussian-process approximations}, 2020.
\newblock URL \url{https://CRAN.R-project.org/package=GPvecchia}.
\newblock R package version 0.1.3.

\bibitem[Katzfuss and Guinness(2021)]{katzfuss_general_2021}
Matthias Katzfuss and Joseph Guinness.
\newblock A general framework for {Vecchia} approximations of {Gaussian}
  processes.
\newblock \emph{Statistical Science}, 36\penalty0 (1):\penalty0 124--141, 2021.
\newblock ISSN 0883-4237, 2168-8745.
\newblock \doi{10.1214/19-STS755}.
\newblock URL \url{http://projecteuclid.org/euclid.ss/1608541222}.

\bibitem[Rubin(1987)]{rubin_multiple_1987}
Donald~B. Rubin, editor.
\newblock \emph{Multiple {Imputation} for {Nonresponse} in {Surveys}}.
\newblock Wiley {Series} in {Probability} and {Statistics}. John Wiley \& Sons,
  Inc., Hoboken, NJ, USA, 1987.
\newblock ISBN 978-0-470-31669-6 978-0-471-08705-2.
\newblock \doi{10.1002/9780470316696}.
\newblock URL \url{http://doi.wiley.com/10.1002/9780470316696}.

\bibitem[Vernet et~al.(2019)Vernet, Geibert, Hoppema, Brown, Haas, Hellmer,
  Jokat, Jullion, Mazloff, Bakker, Brearley, Croot, Hattermann, Hauck,
  Hillenbrand, Hoppe, Huhn, Koch, and Lechtenfeld]{vernet_weddell_2019}
M.~Vernet, W.~Geibert, M.~Hoppema, P.~J. Brown, C.~Haas, H.~H. Hellmer,
  W.~Jokat, L.~Jullion, M.~Mazloff, D.~C.~E. Bakker, J.~A. Brearley, P.~Croot,
  T.~Hattermann, J.~Hauck, C.-D. Hillenbrand, C.~J.~M. Hoppe, O.~Huhn, B.~P.
  Koch, and A.~Lechtenfeld, O. J. ...~Verdy.
\newblock The {Weddell} {Gyre}, {Southern} {Ocean}: {present} {knowledge} and
  {future} {challenges}.
\newblock \emph{Reviews of Geophysics}, 57\penalty0 (3):\penalty0 623--708,
  2019.
\newblock ISSN 1944-9208.
\newblock \doi{10.1029/2018RG000604}.
\newblock URL
  \url{https://agupubs.onlinelibrary.wiley.com/doi/abs/10.1029/2018RG000604}.

\bibitem[Riser et~al.(2018)Riser, Swift, and Drucker]{riser_profiling_2018}
Stephen~C. Riser, Dana Swift, and Robert Drucker.
\newblock Profiling {floats} in {SOCCOM}: {technical} {capabilities} for
  {studying} the {Southern} {Ocean}.
\newblock \emph{Journal of Geophysical Research: Oceans}, 123\penalty0
  (6):\penalty0 4055--4073, 2018.
\newblock ISSN 2169-9291.
\newblock \doi{https://doi.org/10.1002/2017JC013419}.

\bibitem[Cervone and Pillai(2015)]{cervone_gaussian_2015}
Daniel Cervone and Natesh~S. Pillai.
\newblock Gaussian {process} {regression} with {location} {errors}.
\newblock \emph{arXiv:1506.08256 [math, stat]}, 2015.
\newblock URL \url{http://arxiv.org/abs/1506.08256}.
\newblock arXiv: 1506.08256.

\bibitem[Campbell et~al.(2019)Campbell, Wilson, Moore, Riser, Brayton, Mazloff,
  and Talley]{campbell_antarctic_2019}
Ethan~C. Campbell, Earle~A. Wilson, G.~W.~Kent Moore, Stephen~C. Riser,
  Casey~E. Brayton, Matthew~R. Mazloff, and Lynne~D. Talley.
\newblock Antarctic offshore polynyas linked to {Southern} {Hemisphere} climate
  anomalies.
\newblock \emph{Nature}, 570\penalty0 (7761):\penalty0 319--325, 2019.
\newblock ISSN 1476-4687.
\newblock \doi{10.1038/s41586-019-1294-0}.
\newblock URL \url{http://www.nature.com/articles/s41586-019-1294-0}.

\end{thebibliography}

\newpage
\appendix
\section{Additional Figures}


\begin{figure}[h]
\begin{center}
\includegraphics[width=0.4\linewidth]{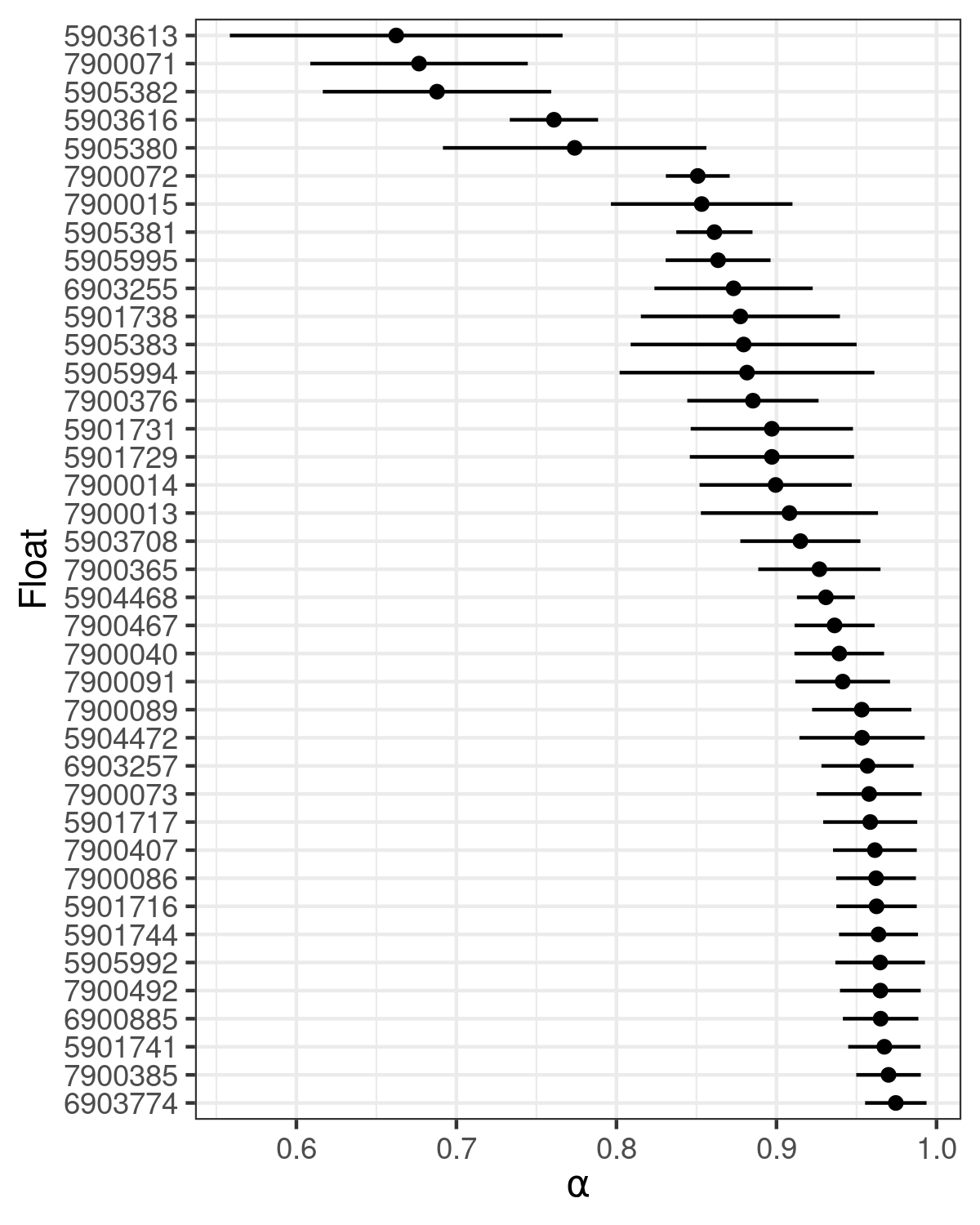}(A)
\includegraphics[width=0.4\linewidth]{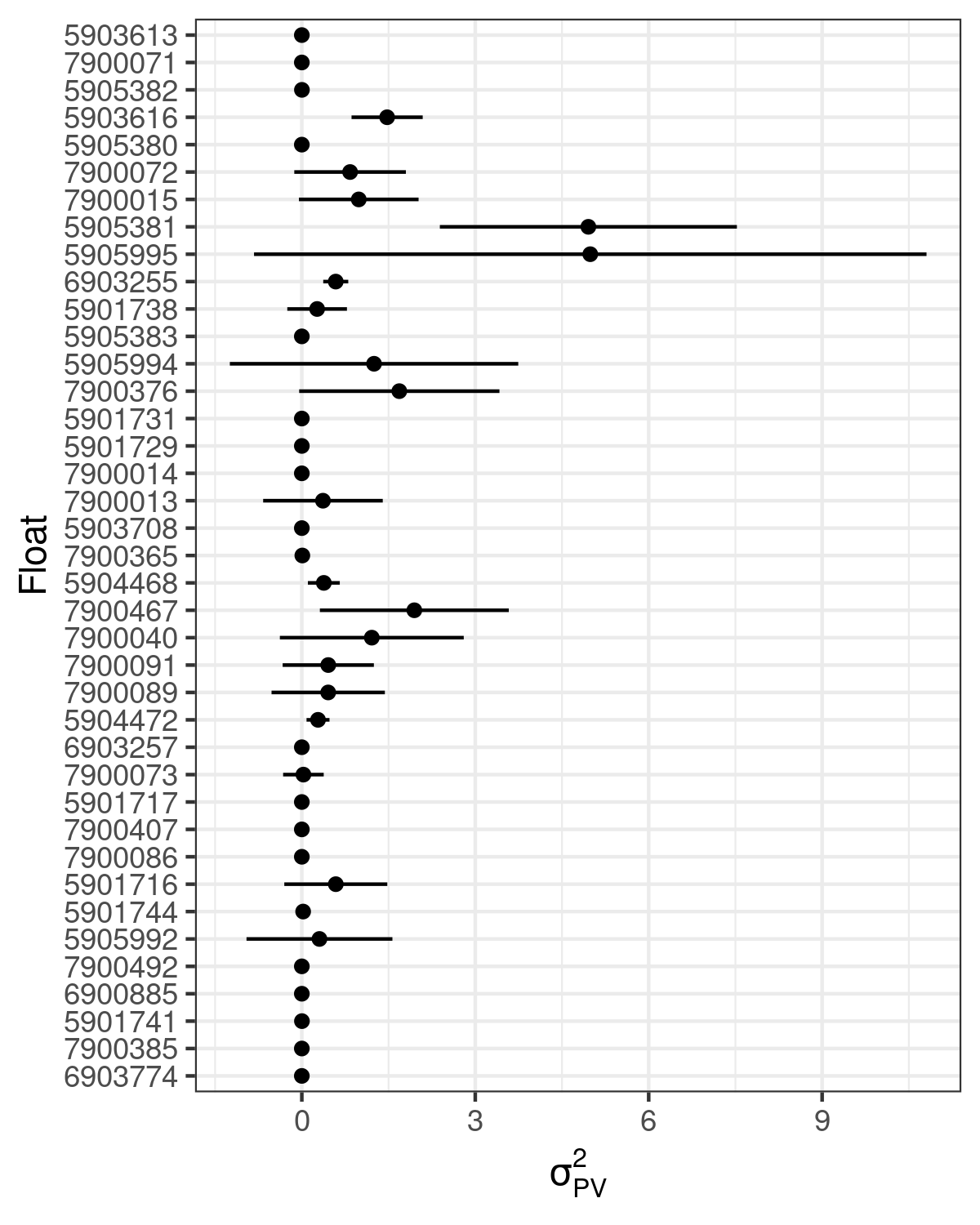}(B)
\includegraphics[width=0.4\linewidth]{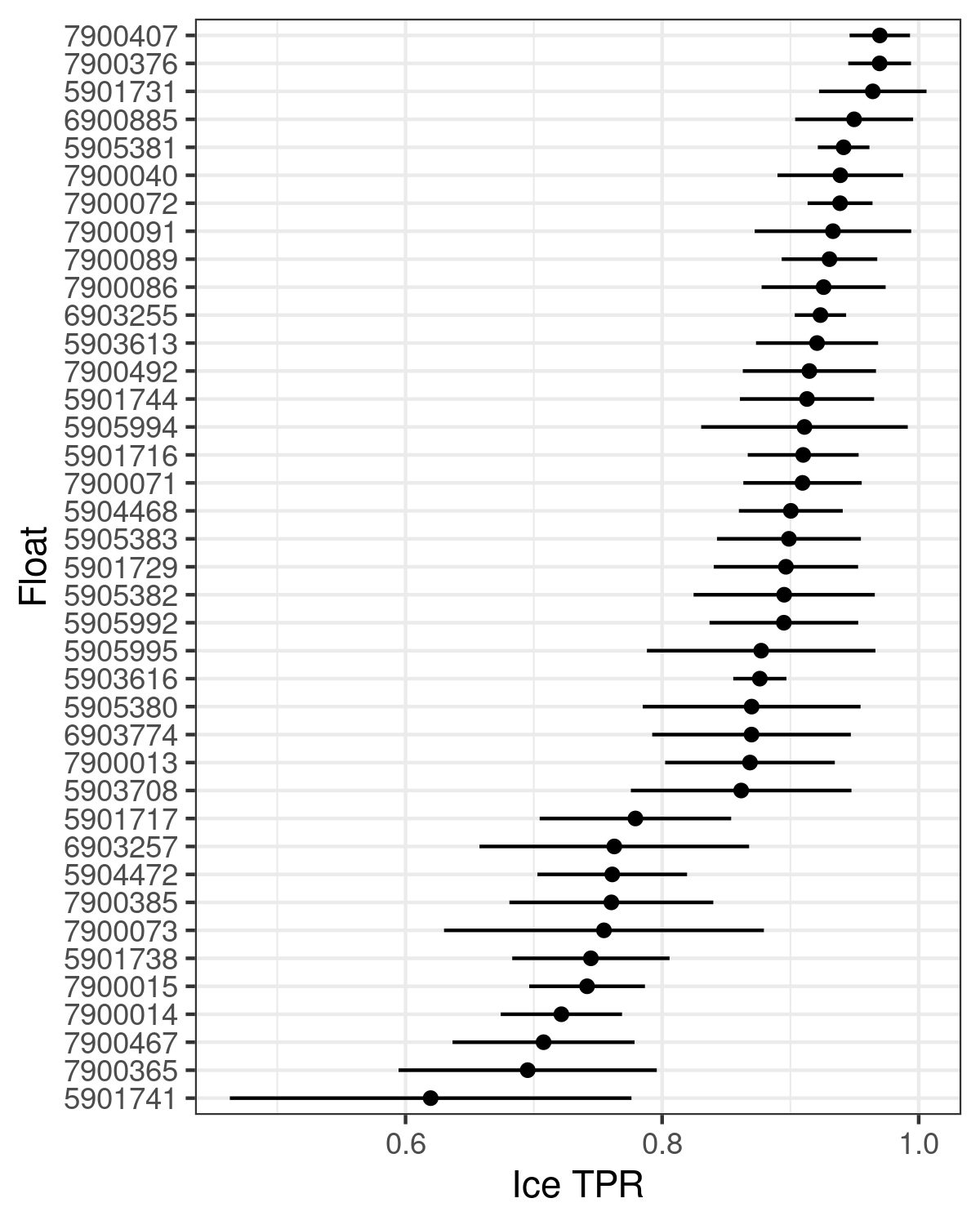}(C)
\includegraphics[width=0.4\linewidth]{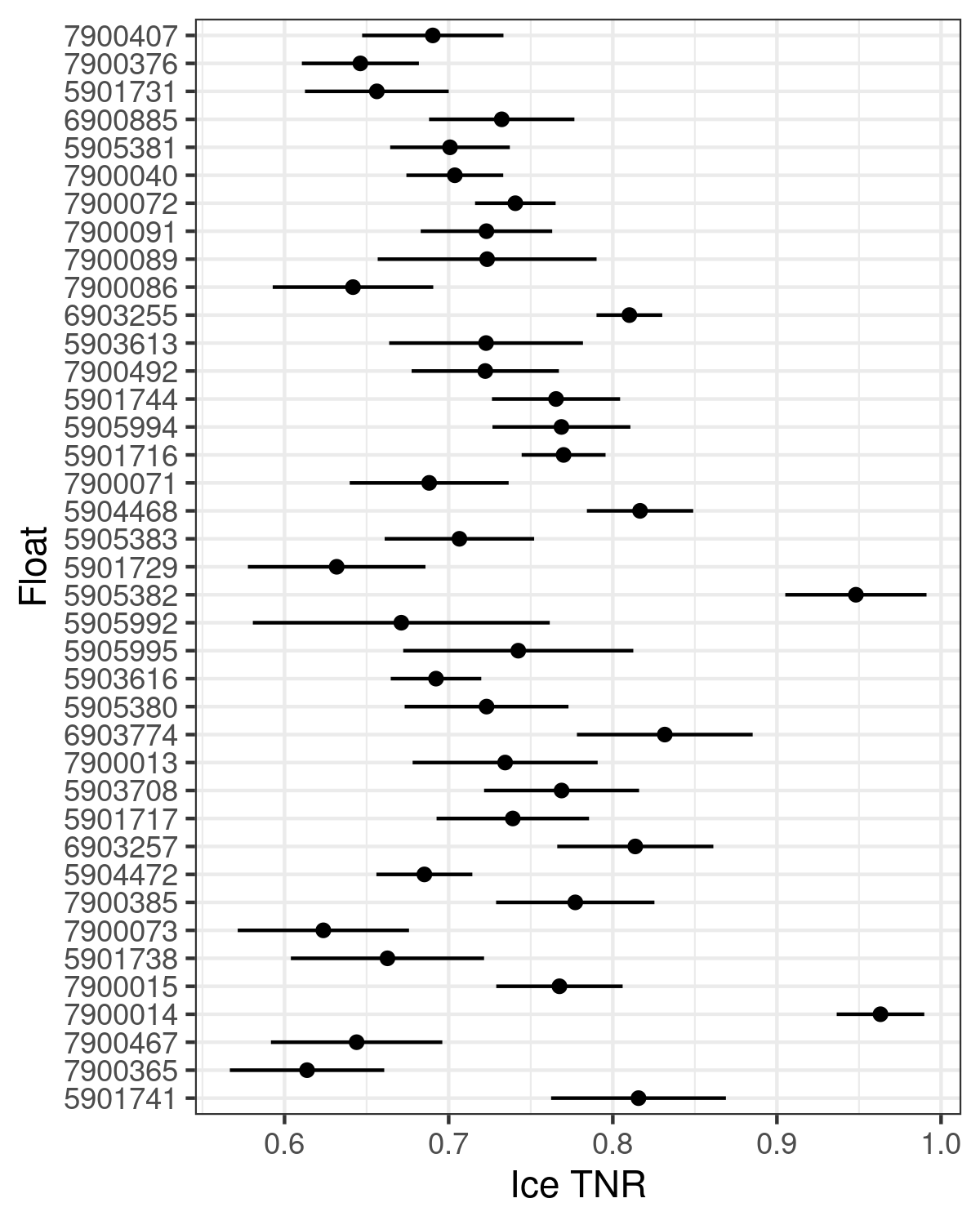}(D)
  \caption{
    Posterior distributions for select parameters across the floats used for spatiotemporal function estimation. 
    The black dot indicates the posterior mean while the black line indicates one standard deviation. 
    The selected parameters are: (A) the autoregressive coefficient $\alpha$; (B) the potential vorticity variance $\sigma_{PV}^2$; (C) the true positive rate (TPR) for the ice-detection algorithm; and (D) the true negative rate (TNR) for the ice detection algorithm.
 } \label{fig:parameter_posteriors}
\end{center}
\end{figure}

\begin{figure}[t]
  \centering
  \begin{minipage}{0.5\linewidth}
\includegraphics[width=\linewidth]{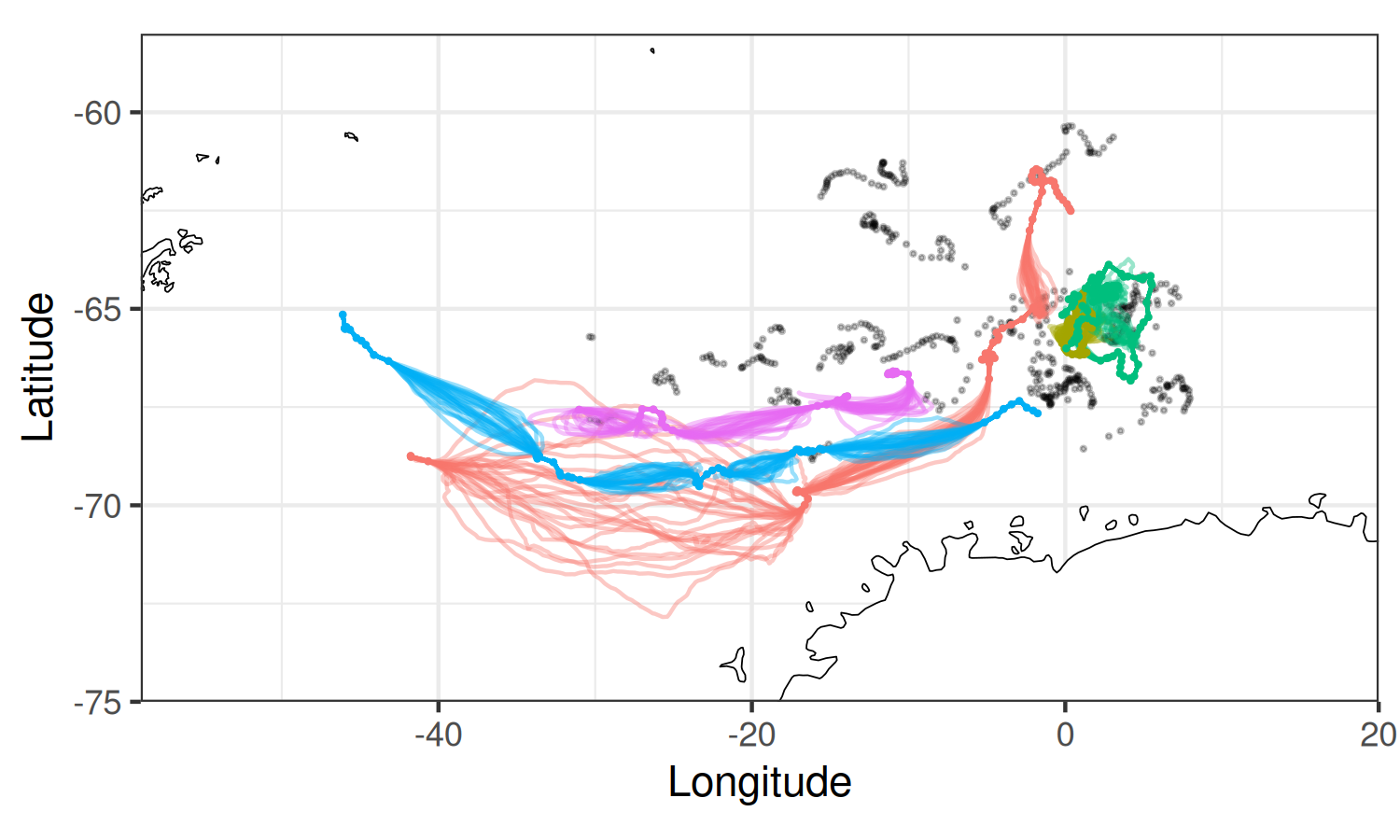}
 (A)
\end{minipage}
\begin{minipage}{0.29\linewidth}
\includegraphics[width=\linewidth]{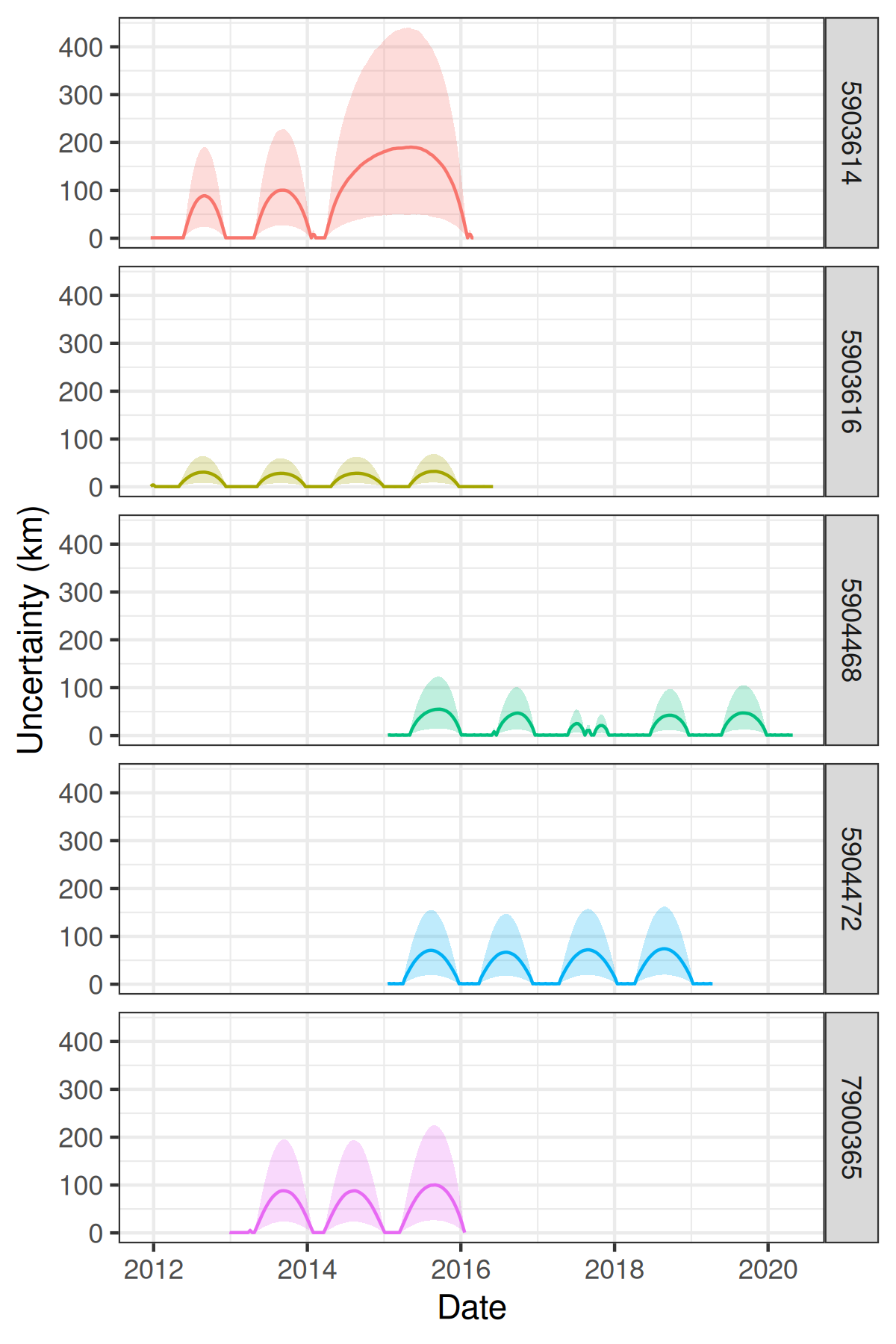}
(B)
\end{minipage} 
  \caption{A view of selected Argo floats that had at least one observation in 2015, which were used in constructing spatial estimates. In subfigure (A), each point represents a GPS observation, while certain floats' predicted trajectories are plotted in color. In subfigure (B), the median uncertainty is plotted over time, with the shaded area representing the interval between the 5\% and 95\% quantiles. Float 5903014 had missing locations for nearly two consecutive years (2014-2015), while float 5904472 received GPS positioning twice during winter in late 2017. More information about the latter event is available in \cite{campbell_antarctic_2019}.} \label{fig:all_floats}
\end{figure}

\begin{figure}[ht]
  \centering
\includegraphics[width=0.8\linewidth]  {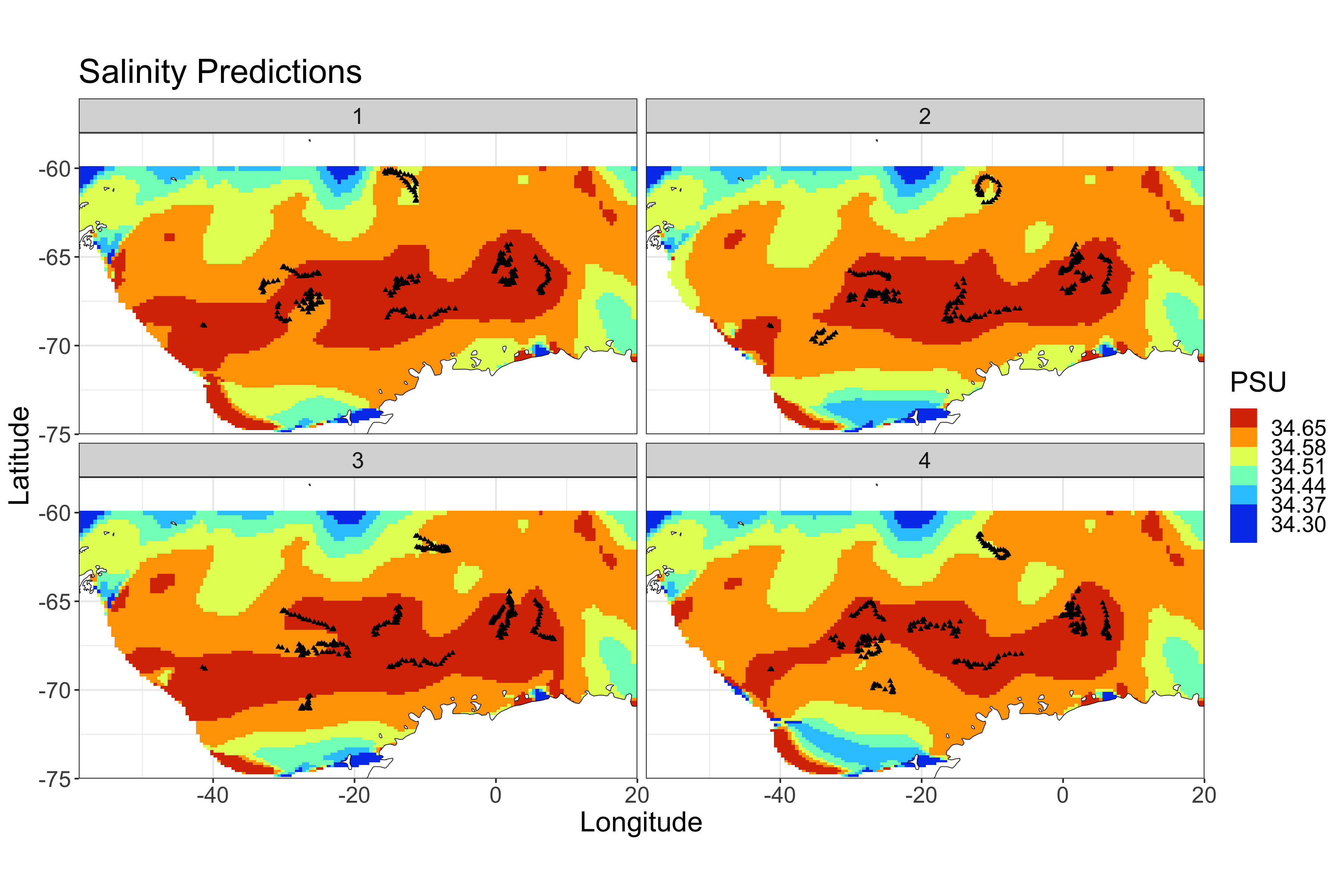}
  \caption{Mean salinity estimates in practical salinity units (PSU) on August 1, 2015 taken on four different samples of locations. The black dots show the ArgoSSM sample of missing locations that was used to construct the estimate.}
  \label{fig:samples_psal}
\end{figure}

\begin{figure}[ht]
  \centering
\begin{minipage}{0.45\linewidth}
  \includegraphics[width=0.8\linewidth]{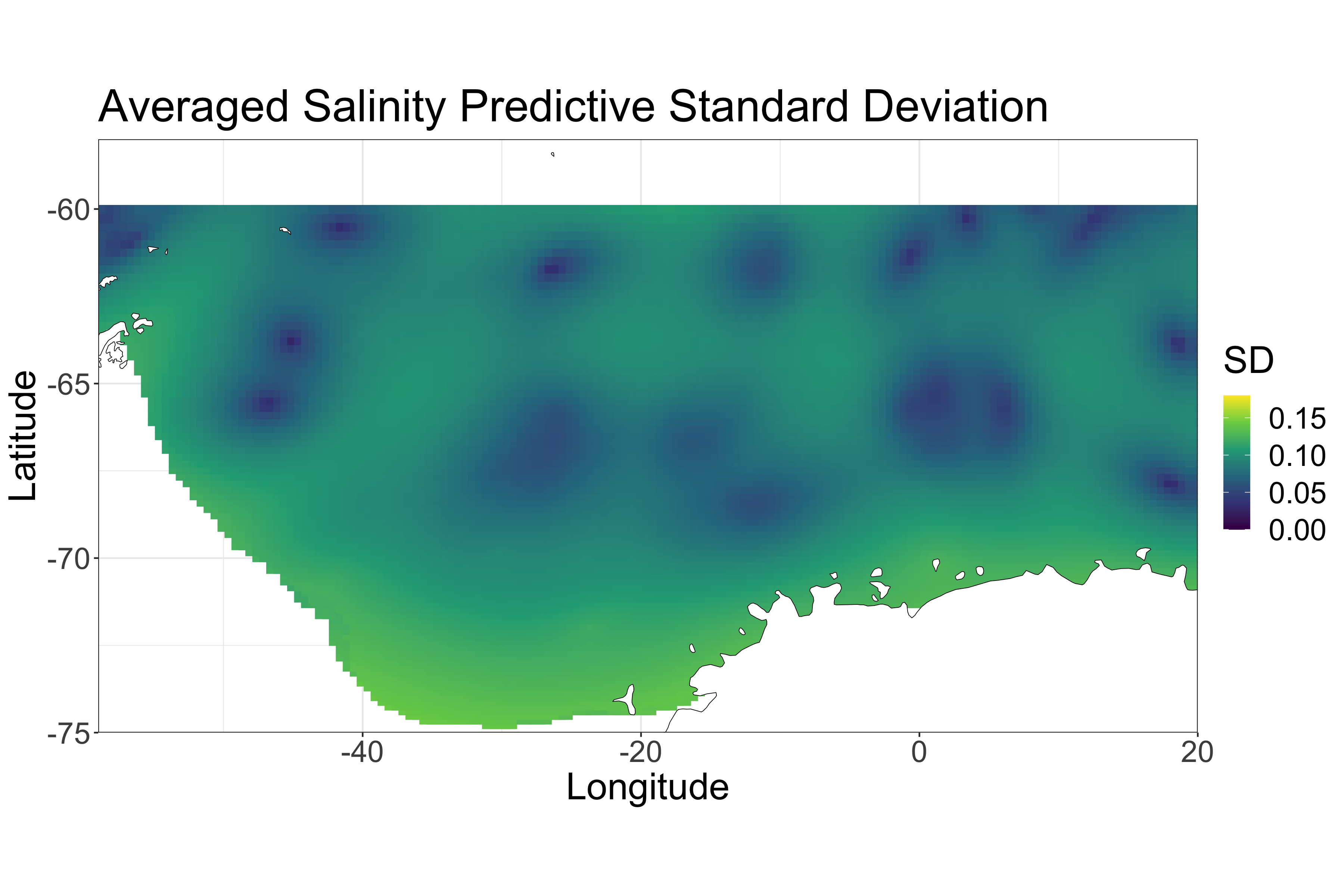}

  (A)
\end{minipage}
\begin{minipage}{0.45\linewidth}
  \includegraphics[width=0.8\linewidth]  {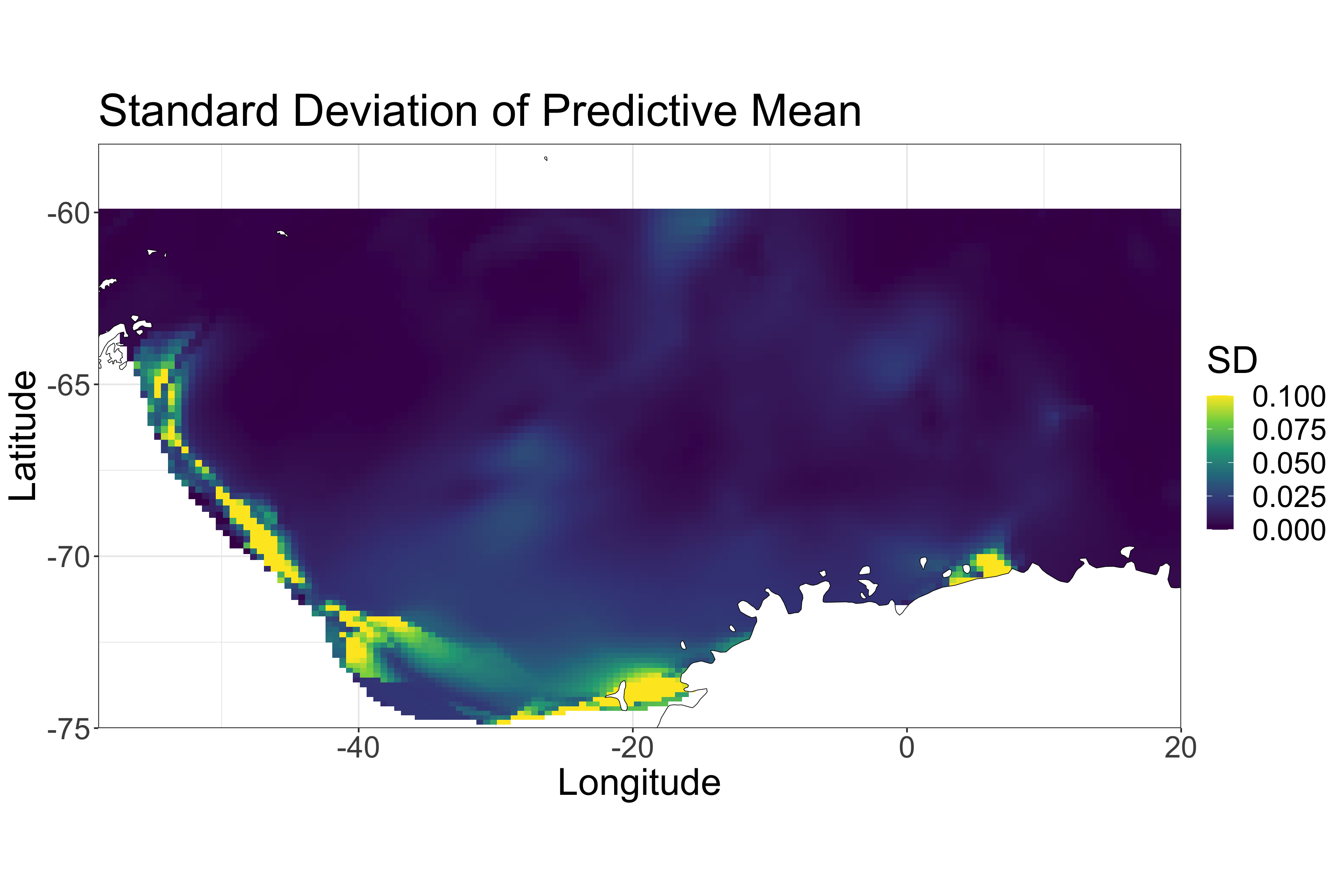}

  (B)
\end{minipage}

\begin{minipage}{0.45\linewidth}
  \includegraphics[width=0.8\linewidth]  {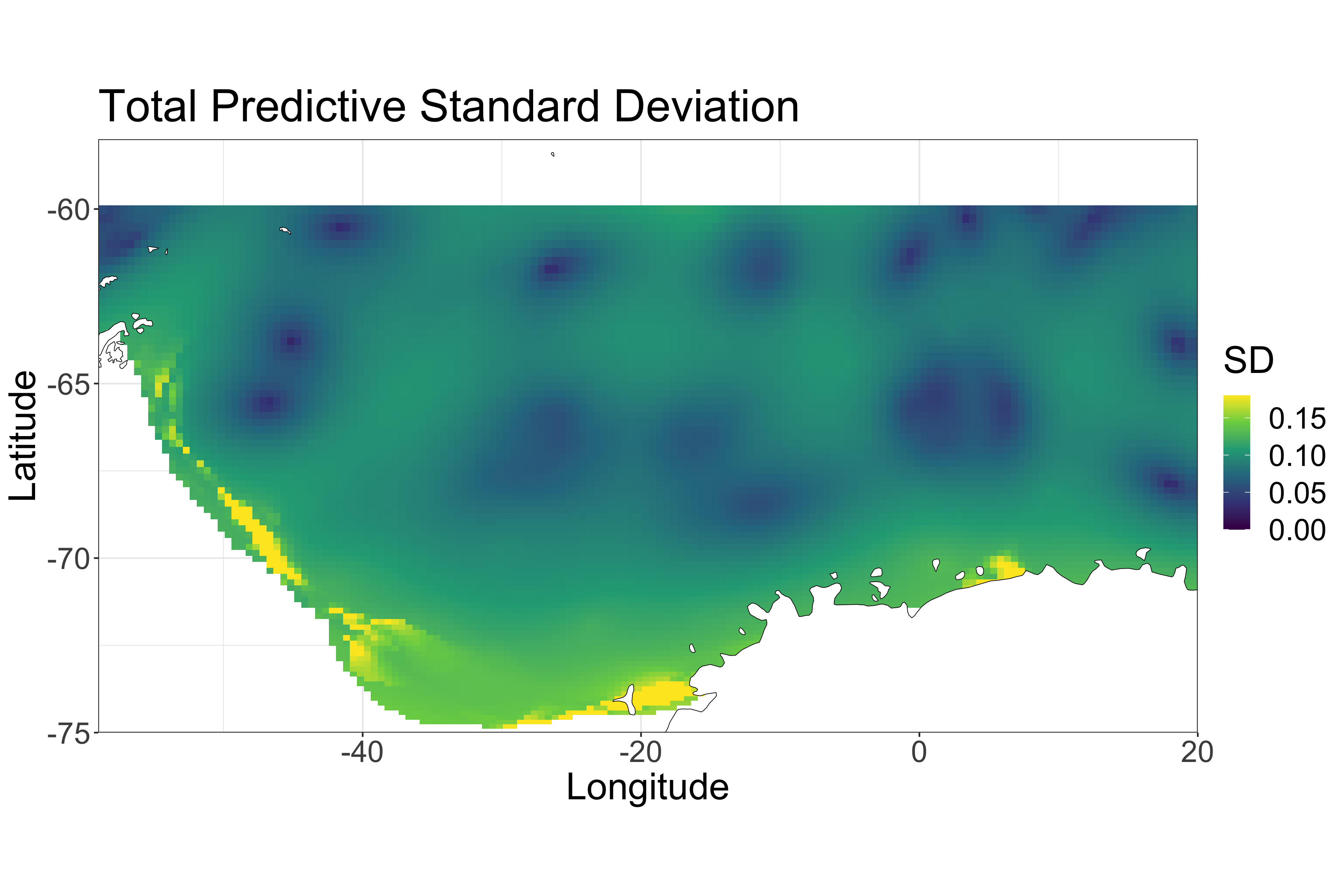}

  (C)
\end{minipage}

  \caption{Spatial uncertainty in salinity estimated on August 1, 2015. (A) shows the average standard deviation conditional on locations. (B) shows the standard deviation of mean temperature estimates between samples of locations. (C) shows the total standard deviation accounting for both (A) and (B).}
  \label{fig:est_psal}
\end{figure}

\begin{figure}
\centering
\includegraphics[scale = .13]{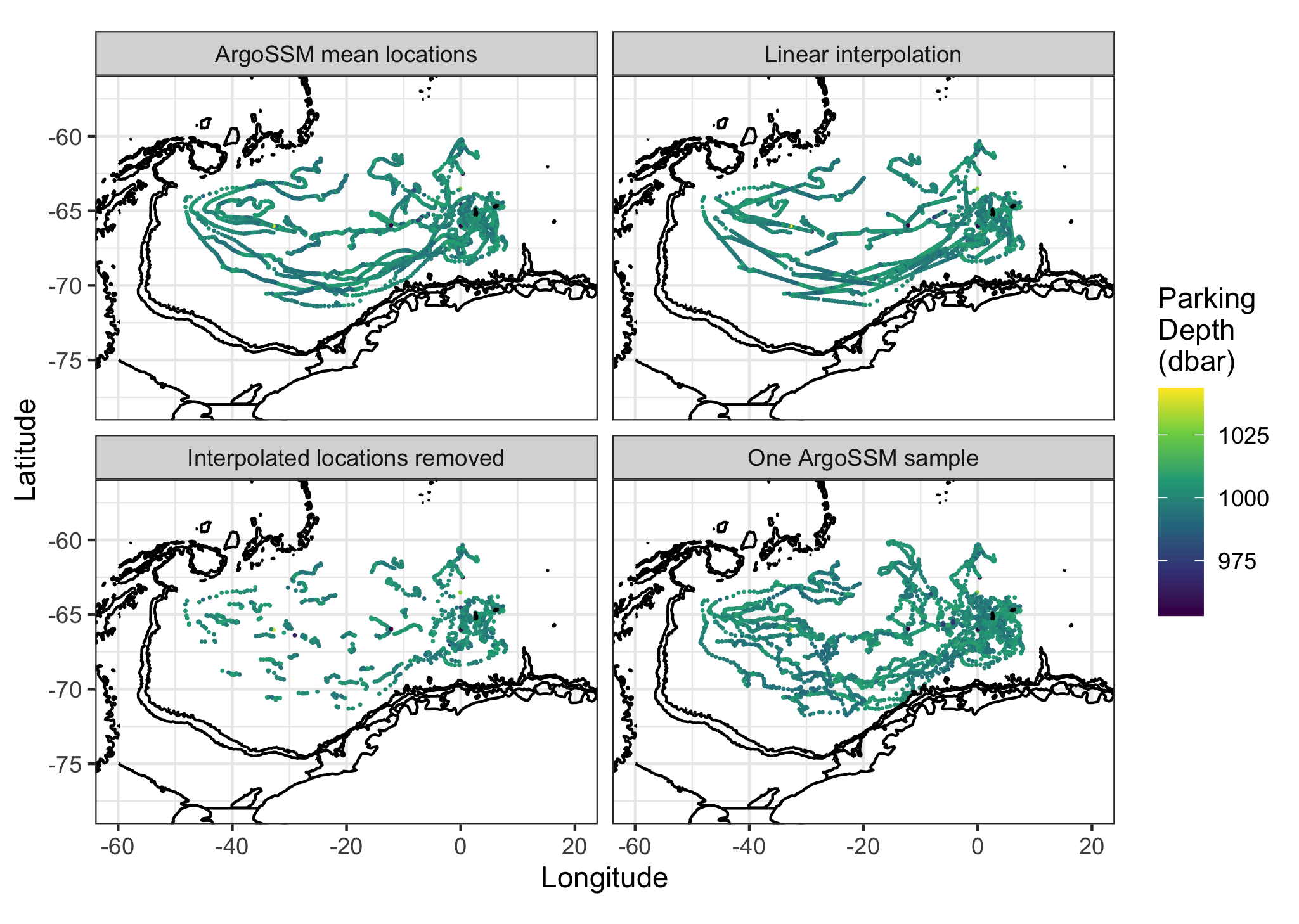}
\caption{Float trajectories used for velocity estimation, with (from top left, clockwise) ArgoSSM mean, linearly interpolated values, a sample from ArgoSSM, and with missing trajectories removed.}\label{fig:veldata}
\end{figure}


\end{document}